\documentclass[twocolumn]{aastex631}

\let\tablenum\relax
\usepackage{siunitx}
\usepackage{amsmath}
\usepackage{multirow}
\usepackage[utf8]{inputenc}
\usepackage{array}

\newcommand{\Hb}{\sc H$\beta$}
\newcommand{\Oiii}{\sc [Oiii]}
\newcommand{\OOiii}{\sc{[Oiii]}$\lambda$5007}
\newcommand{\Ha}{\sc H$\alpha$}
\newcommand{\Hg}{\sc H$\gamma$}
\newcommand{\OOii}{\sc{[Oii]}$\lambda\lambda$3727,29}
\begin{document}

\title{The Physical Origin of Extreme Emission Line Galaxies at High redshifts:\\
Strong {\Oiii} Emission Lines Produced by Obscured AGNs
}

\author[0000-0002-9888-6895]{Chenghao Zhu}
\affiliation{Institute for Cosmic Ray Research, The University of Tokyo, 5-1-5 Kashiwanoha, Kashiwa, Chiba 277-8582, Japan}
\affiliation{Department of Physics, Graduate School of Science, The University of Tokyo, 7-3-1 Hongo, Bunkyo, Tokyo 113-0033, Japan}

\author[0000-0002-6047-430X]{Yuichi Harikane} 
\affiliation{Institute for Cosmic Ray Research, The University of Tokyo, 5-1-5 Kashiwanoha, Kashiwa, Chiba 277-8582, Japan}

\author[0000-0002-1049-6658]{Masami Ouchi}
\affiliation{National Astronomical Observatory of Japan, 2-21-1 Osawa, Mitaka, Tokyo 181-8588, Japan}
\affiliation{Institute for Cosmic Ray Research, The University of Tokyo, 5-1-5 Kashiwanoha, Kashiwa, Chiba 277-8582, Japan}
\affiliation{Department of Astronomical Science, SOKENDAI (The Graduate University for Advanced Studies), Osawa 2-21-1, Mitaka, Tokyo, 181-8588, Japan}
\affiliation{Kavli Institute for the Physics and Mathematics of the Universe (WPI), University of Tokyo, Kashiwa, Chiba 277-8583, Japan}

\author[0000-0001-9011-7605]{Yoshiaki Ono}
\affiliation{Institute for Cosmic Ray Research, The University of Tokyo, 5-1-5 Kashiwanoha, Kashiwa, Chiba 277-8582, Japan}

\author[0000-0003-3228-7264]{Masato Onodera}
\affiliation{Department of Astronomical Science, SOKENDAI (The Graduate University for Advanced Studies), Osawa 2-21-1, Mitaka, Tokyo, 181-8588, Japan}
\affiliation{Subaru Telescope,  National Astronomical Observatory of Japan, 650 North Aohoku Place, Hilo, HI, 96720, USA}

\author[0000-0002-2185-5679]{Shenli Tang}
\affiliation{School of Physics \& Astronomy, University of Southampton, Highfield Campus, Southampton SO17 1BJ, UK}

\author[0000-0001-7730-8634]{Yuki Isobe}
\affiliation{Waseda Research Institute for Science and Engineering, Faculty of Science and Engineering, Waseda University, 3-4-1, Okubo, Shinjuku, Tokyo 169-8555, Japan}

\author[0000-0001-5063-0340]{Yoshiki Matsuoka}
\affiliation{Research Center for Space and Cosmic Evolution, Ehime University,
Matsuyama, Ehime 790-8577, Japan}

\author[0000-0002-3866-9645]{Toshihiro Kawaguchi}
\affiliation{Department of Economics, Management and Information Science, Onomichi City University, Hisayamada 1600-2, Onomichi, Hiroshima 722-8506, Japan}

\author{Hiroya Umeda}
\affiliation{Institute for Cosmic Ray Research, The University of Tokyo, 5-1-5 Kashiwanoha, Kashiwa, Chiba 277-8582, Japan}
\affiliation{Department of Physics, Graduate School of Science, The University of Tokyo, 7-3-1 Hongo, Bunkyo, Tokyo 113-0033, Japan}

\author[0000-0003-2965-5070]{Kimihiko Nakajima}
\affiliation{National Astronomical Observatory of Japan, 2-21-1 Osawa, Mitaka, Tokyo 181-8588, Japan}

\author[0000-0002-2725-302X]{Yongming Liang}
\affiliation{Institute for Cosmic Ray Research, The University of Tokyo, 5-1-5 Kashiwanoha, Kashiwa, Chiba 277-8582, Japan}

\author[0000-0002-5768-8235]{Yi Xu}
\affiliation{Institute for Cosmic Ray Research, The University of Tokyo, 5-1-5 Kashiwanoha, Kashiwa, Chiba 277-8582, Japan}
\affiliation{Department of Astronomy, Graduate School of Science, The University of Tokyo, 7-3-1 Hongo, Bunkyo, Tokyo 113-0033, Japan}

\author[0000-0003-3817-8739]{Yechi Zhang}
\affiliation{National Astronomical Observatory of Japan, 2-21-1 Osawa, Mitaka, Tokyo 181-8588, Japan}

\author[0000-0002-1199-6523]{Dongsheng Sun}
\affiliation{Institute for Cosmic Ray Research, The University of Tokyo, 5-1-5 Kashiwanoha, Kashiwa, Chiba 277-8582, Japan}
\affiliation{Department of Astronomy, Graduate School of Science, The University of Tokyo, 7-3-1 Hongo, Bunkyo, Tokyo 113-0033, Japan}
\author[0000-0002-2597-2231]{Kazuhiro Shimasaku}
\affiliation{Department of Astronomy, University of Tokyo, Bunkyo, Tokyo 113-0033, Japan}

\author{Jenny Greene}
\affiliation{Department of Astrophysical Sciences, Princeton University, Princeton, NJ 08544, USA}

\author[0000-0002-4923-3281]{Kazushi Iwasawa}
\affiliation{ICCUB, Universitat de Barcelona (IEEC-UB), Mart\'{i} i Franqu\`{e}s, 1,08028 Barcelona, Spain}
\affiliation{ICREA, Pg Llu\'is Companys 23, 08010 Barcelona, Spain}

\author{Kotaro Kohno}
\affiliation{Institute of Astronomy, University of Tokyo, Mitaka, Tokyo 181-0015, Japan}

\author{Tohru Nagao}
\affiliation{Research Center for Space and Cosmic Evolution, Ehime University, 2-5 Bunkyo-cho, Matsuyama, Ehime 790-8577, Japan}
\affiliation{Amanogawa Galaxy Astronomy Research Center, Kagoshima University, 1-21-35 Korimoto, Kagoshima 890-0065, Japan}

\author{Andreas Schulze}
\affiliation{National Astronomical Observatory of Japan, 2-21-1 Osawa, Mitaka, Tokyo 181-8588, Japan}

\author{Takatoshi Shibuya}
\affiliation{Kitami Institute of Technology, Kitami, Hokkaido 090-8507, Japan}

\author[0000-0001-8587-1582]{Miftahul Hilmi}
\affiliation{School of Physics, the University of Melbourne, Parkville, VIC 3010, Australia}
\affiliation{ARC Centre of Excellence for All Sky Astrophysics in 3 Dimensions (ASTRO 3D), Australia}

\author{Malte Schramm}
\affiliation{National Astronomical Observatory of Japan, 2-21-1 Osawa, Mitaka, Tokyo 181-8588, Japan}




\begin{abstract}
We present deep Subaru/FOCAS spectra for two extreme emission line galaxies (EELGs) at $z\sim 1$ with strong {\sc[Oiii]}$\lambda$5007 emission lines, exhibiting equivalent widths (EWs) of $2905^{+946}_{-578}$ \AA\ and $2000^{+188}_{-159}$ \AA, comparable to those of EELGs at high redshifts that are now routinely identified with JWST spectroscopy. Adding a similarly large {\Oiii} EW ($2508^{+1487}_{-689}$ \AA) EELG found at $z\sim 2$ in the JWST CEERS survey to our sample, we explore for the physical origins of the large {\Oiii} EWs of these three galaxies with the Subaru spectra and various public data including JWST/NIRSpec, NIRCam, and MIRI data. While there are no clear signatures of AGN identified by the optical line diagnostics, we find that two out of two galaxies covered by the MIRI data show strong near-infrared excess in the spectral energy distributions (SEDs) indicating obscured AGN. 
Because none of the three galaxies show clear broad H$\beta$ lines, the upper limits on the flux ratios of broad-H$\beta$ to {\Oiii} lines are small, $\lesssim 0.15$ that are comparable with Seyfert $1.8-2.0$ galaxies. 
We conduct \texttt{Cloudy} modeling with the stellar and AGN incident spectra, allowing a wide range of parameters including metallicities and ionization parameters. We find that the large {\Oiii} EWs are not self-consistently reproduced by the spectra of stars or unobscured AGN, but obscured AGN that efficiently produces O$^{++}$ ionizing photons with weak nuclear and stellar continua that are consistent with the SED shapes.



%

\end{abstract}



\section{Introduction}
\label{sec:intro}
Nebular spectra can provide information on understanding the galaxies' properties. Emission lines can be used as a probe to trace the incident UV radiation (e.g. \citealt{kewleyUnderstandingGalaxyEvolution2019}; \citealt{oeyCalibrationNebularEmission2000a}; \citealt{isobeEMPRESSIVExtremely2022}). 
Particularly, the forbidden lines {\Oiii$\lambda\lambda$4959,5007} ({\Oiii} doublets) play a pivotal role.
The high excitation lines are often driven by ionizing photons produced in the massive and short-lived O and B stars or the active galactic nuclei (AGN), whereas they are surrounded by the rest-optical continuum mainly contributed by the less massive and longer-lived stars. 

Extreme emission line galaxies (EELGs) are characterized by their notably strong line emissions in comparison to their stellar continuum, resulting in unusually high emission line equivalent widths (EWs). Over the last decade, the EELGs have been studied in detail at very low redshift, especially those galaxies identified by their extremely large EWs of {\OOiii} ({\Oiii} emitters; e.g. the ``green pea" population, \citealt{cardamoneGalaxyZooGreen2009}; \citealt{izotovGREENPEAGALAXIES2011}; \citealt{jaskotORIGINOPTICALDEPTH2013}; and the ``blueberry" population \citealt{yangBlueberryGalaxiesLowest2017}). The typical {\Oiii} EW is 20 {\AA} in the Sloan Digital Sky Survey (SDSS; e.g. \citealt{alamELEVENTHTWELFTHDATA2015}), and only $<1\%$ of SDSS galaxies exhibit {\Oiii} EWs $\gtrsim$ 1000 {\AA}. Most of the EELGs are considered to be undergoing strong star formation (\citealt{izotovGREENPEAGALAXIES2011}; \citealt{boyettOIIILambda5007Equivalent2022}). In comparison, several studies (e.g. \citealt{baskinWhatControlsIii2005}; \citealt{caccianigaRelationshipIiiL50072011}; \citealt{mullaneyNarrowlineRegionGas2013}) show that the narrow line EWs of AGN can also have a large value of EW({\Oiii}) $\sim2000${\AA}, suggesting that not all EELGs should be simply attributed to star-forming galaxies.

The launch of the James Webb Space Telescope provides abundant unprecedented data in terms of both spectra and images \citep{rigbySciencePerformanceJWST2023}. It enables sensitive near-infrared spectroscopy out to $5.2\mu$m with NIRSpec \citep{jakobsenNearInfraredSpectrographNIRSpec2022} permitting direct measurement of the EWs of {\OOiii} out to high redshift ($z < 9.5$). The recent researches indicate that in the early universe, the EELGs are significantly more abundant (e.g. \citealt{mattheeEIGERIIFirst2023}; \citealt{sunFirstSampleHa2023}; \citealt{boyettExtremeEmissionLine2024}; \citealt{toppingMetalpoorStarFormation2024}). Plus, the MIRI \citep{wrightMidinfraredInstrumentJWST2023} on JWST provides nine photometric bands from 5 to 26 $\mu$m, which are about 10-100 times more sensitive than the Spitzer mission. It allows us to further inspect the mid-infrared properties of low-z galaxies, such as the hot dust emission.

The physical origins of EELGs determined by optical line diagnostics 
remain more uncertain than expected. 
According to the diagnostics of Baldwin, Phillips, and Terlevich (BPT; \citealt{baldwinClassificationParametersEmissionline1981}) diagram, most of the EELGs are considered to be undergoing strong star-forming activities, while a small fraction are AGNs (e.g. `Galaxy Zoo': AGN fraction is 17\%; \citealt{cardamoneGalaxyZooGreen2009}). 
%
It is noteworthy that recent reports suggest that the low-metallicity AGNs with $Z < Z_\odot$ are located in the same region as star-forming galaxies in the classical BPT diagram (e.g. \citealt{izotovActiveGalacticNuclei2008}; \citealt{harikaneJWSTNIRSpecFirst2023}; \citealt{maiolinoJADESDiversePopulation2023}; \citealt{chisholmNeEmissionFaint2024}; \citealt{Yao_2024}). In fact, the majority of local EELGs feature low metallicity ($Z\lesssim0.2Z_\odot$; \citealt{boyettOIIILambda5007Equivalent2022}), indicating that the BPT diagram is not sufficient to determine the physical origins of low-metallicity EELGs.
%
%

Particularly, in the Subaru High-z Exploration of Low-luminosity Quasars (SHELLQs) survey, \citet{matsuokaSubaruHighExploration2018} spectroscopically identified several {\Oiii} emitters at $z \sim 0.8$. Notably, two of these emitters are reported to have unprecedentedly large EWs of {\Oiii}, exceeding 4000{\AA}. However, such high EWs of {\OOiii} are difficult to explain with classic stellar population models that typically predict lower EWs ($\lesssim3000${\AA}; \citealt{inoueRestframeUltraviolettoopticalSpectral2011}). Therefore, we conducted deeper follow-up observations, aiming to {examine the previous measurement and} uncover the underlying mechanisms driving these extraordinary emission characteristics. Furthermore, we search for similarly extreme EW objects in the JWST programs.

This paper is structured as follows. Section \ref{sec:data} provides the details of the new observation and the JWST dataset we use. We outline the basic information about objects, data reduction, and selection criteria. Section \ref{sec:result} presents the basic physical properties implied by the spectroscopic and photometric measurement. Section \ref{sec:discuss} discusses the physical origins of {\Oiii} emitters and implications for our findings. Throughout this paper, we assume the standard $\rm \Lambda CDM$ model with the cosmological parameters from Planck 2018 \citep{planckcollaborationPlanck2018Results2020}: $\Omega_m=0.3111$, $\Omega_\Lambda=0.6899$, $\Omega_b=0.0489$, $h=0.6766$, and $\sigma_8=0.8102$. We measure the EWs in the rest frame. We adopt the solar abundance from \citet{asplundChemicalCompositionSun2009} ($Z_\odot = 0.014$).

\section{Sample \& data}\label{sec:data}
\begin{table*}[!t]
    \caption{Source information for our {\sc [Oiii]} emitters}
    \begin{ruledtabular}
    \begin{tabular}{c c c c c}
    \colhead{ID} & \colhead{RA} & \colhead{Dec} & \colhead{redshift} & \colhead{$t_{\rm exp}$(h)\tablenotemark{$\dagger$} } \\\tableline
    J1000+0211 & 10:00:12.46 & +02:11:27.4 & 0.828 & 4.33 \\
    J0845$-$0123 & 08:45:16.54 & $-$01:23:21.6 & 0.728 & 3.67 \\
    CEERS-3506 & 14:20:37.51 & +53:03:35.6 & 2.055 & 1.70
    \end{tabular}
    \end{ruledtabular}
    \tablenotetext{\dagger}{The total exposure time of the spectra used in this work, which covers the {\OOiii} emission line.}
    \label{tab:basicinfo}
\end{table*}
\subsection{Subaru}\label{subsec:subaru}
\subsubsection{Subaru Sample}\label{subsubsec:Ssample}
We select the two most extreme sources from the SHELLQs survey \citep{matsuokaSubaruHighExploration2018}. Their basic properties are summarized in Table \ref{tab:basicinfo}.

\subsubsection{Subaru Observation}\label{subsubsec:FOCASobs}
We conducted new spectroscopic observations with the Faint Object Camera and Spectrograph (FOCAS; \citealt{kashikawaFOCASFaintObject2002}) on the Subaru Telescope. These observations took place over two consecutive nights, 2020 December 24 - 25, during the S20B semester (Proposal ID: S20B0002N; PI: Y. Harikane). We operated FOCAS in its multi-object spectroscopy mode, employing the VPH850 grism in conjunction with the SO58 order-cut filter. This setup covered a wavelength range of $5,800 - 10,000$ \AA. The slit width was set to 0\farcs8, which yielded a spectral resolution of $R \sim 1500$. Integration times were set to 3.67 and 4.33 hours for J0845$-$0123 and J1000+0211, respectively, which are significantly longer than those of the FOCAS observations in \citet{matsuokaSubaruHighExploration2018} ($\sim10$ minutes for each object), who firstly identified these objects. We also utilized VPH650 grism to obtain the spectra in the observed wavelength range of $5300-7700$\AA, to cover the {\OOii} doublet emission lines with a resolution of $R\sim2500$. The total integrated exposure time of observations with VPH650 is 20 minutes for each object.
%
%

\subsubsection{Subaru Data Reduction}\label{subsubsec:Sred}
\begin{figure*}[!thb]
    \plotone{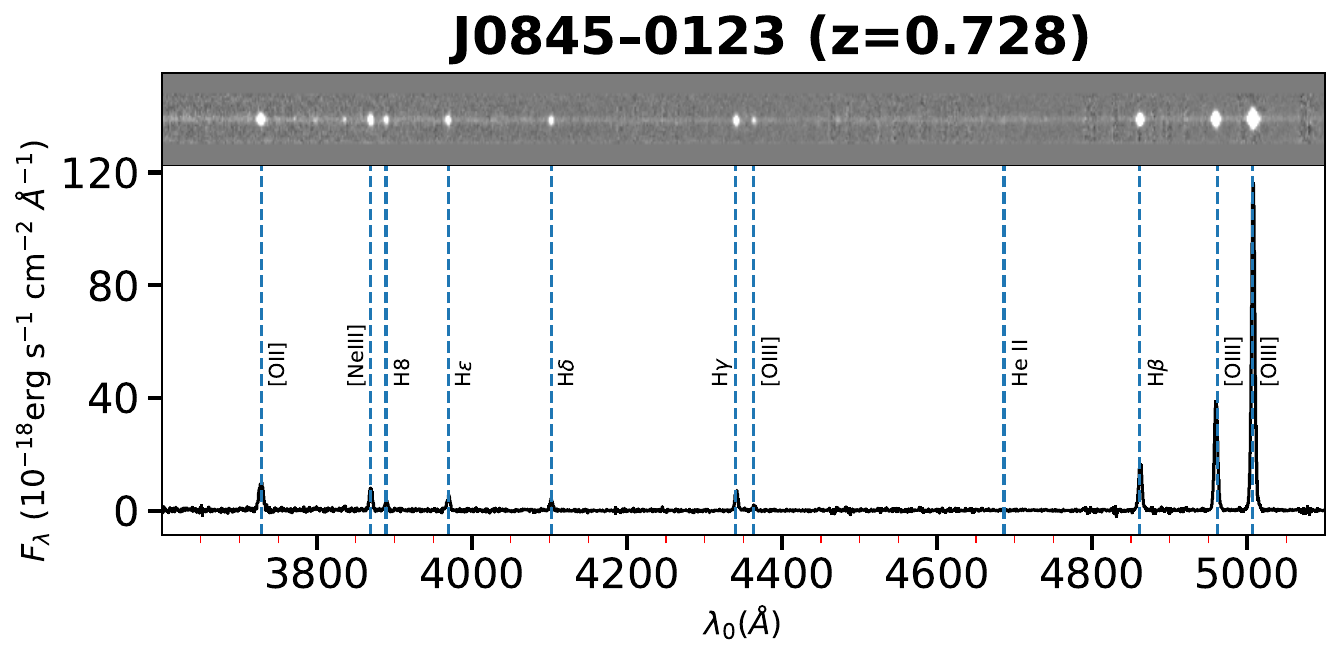}
    \plotone{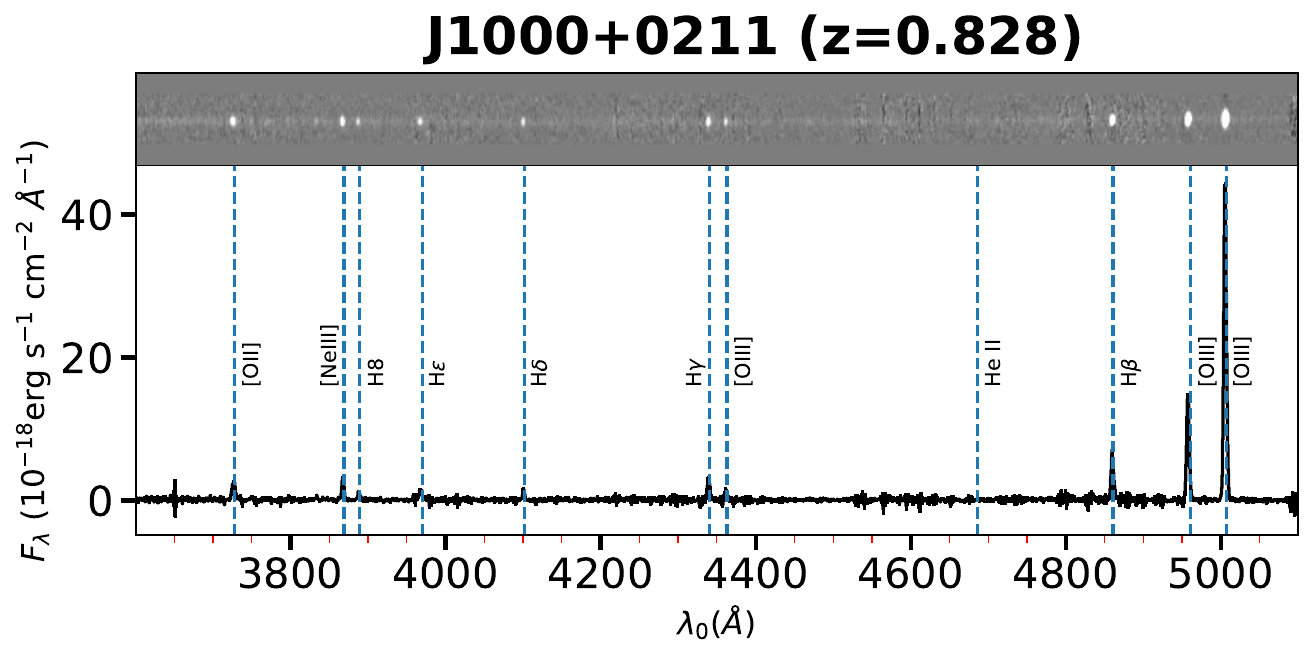}
    \caption{Rest-frame 1D and 2D spectra of J0845$-$0123 (top) and J1000+0211 (bottom). 
    They exhibit several emission lines including the strong {\Hb +\Oiii} doublet emission lines. The faint continuum is detected.}
    \label{fig:subaruspec}
\end{figure*}
In this section, we detail the reduction process of the Subaru data.
Utilizing the Image Reduction and Analysis Facility (IRAF) along with the FOCASRED package from the FOCAS official website, we carry out a series of standard data reduction procedures including bias and overscan subtraction, flat-fielding corrections, and background subtraction. We conduct wavelength calibration by identifying sky emission lines. Flux calibration is achieved with the standard star Feige 34. 
We make error spectra consisting of 
%
Poisson photon noise and read-out noise that is estimated with the CCD overscan regions.
%
Figure \ref{fig:subaruspec} displays the 2D and 1D spectra featuring the strong {\sc H$\beta$} + {\sc [Oiii]} doublets in the rest frame.

\subsection{Keck}
We observed J1000+0211 and J0845$-$0123 with the Multi-Object Spectrometer for Infra-Red Exploration (MOSFIRE) on the Keck I telescope on 2020 January 20 (Proposal ID: S19B0052; PI: Y. Harikane). The spectra are taken with the J-band covering $1.15-1.35$ $\mu$m, targeting the {\Ha}, {\sc [Nii]$\lambda$6484}, and {\sc [Sii]$\lambda\lambda$6717,37} lines redshifted to $z\sim0.8$. The total integration time is 50 minutes for each object. The average seeing size is $\sim 0\farcs6 - 0\farcs9$ for both J1000+0211 and J0845$-$0123. The slit width is 0\farcs7 leading to a spectral resolution of $R\sim3318$.

The data are reduced by using the MOSFIRE data reduction pipeline\footnote{\url{http://code.google.com/p/mosfire}}. This pipeline performs flat fielding, wavelength calibration, sky subtraction, and cosmic ray removal. After combining the spectra, we detect the {\Ha} line in J1000+0211, while the {\Ha} line of J0845-0123 is in the wavelength gap.

\subsection{JWST}\label{subsec:jwst}
%
%
%
We extensively attempt to search 
for sources similar to 
Subaru emitters in the publicly available JWST datasets. 
%

\subsubsection{JWST Sample}\label{subsubsec:jwstsample}
The Cosmic Evolution Early Release Science (CEERS; ERS 1345; PI: S. Finkelstein; \citealt{bagleyCEERSEpochNIRCam2023}; \citealt{finkelsteinCEERSKeyPaper2023}) data were taken with the JWST/NIRSpec Prism covering 0.6--5.3 $\mu$m as well as the medium-resolution filter-grating pairs of F100LP-G140M, F170LP-G235M, and F290LP-G395M covering the wavelength ranges of $1.0-1.6$, $1.7-3.1$, and $2.9-5.1$ $\mu$m, respectively. 

In our work, we use the data reduced by the \texttt{Grizli} \citep{brammer_2023_8370018}, and they are made available through the Cosmic Dawn Center. These data 
can be retrieved from
the DAWN JWST Archive (DJA).\footnote{\url{https://dawn-cph.github.io/dja/}} Details of the reduction of the DJA data are presented in \citet{valentinoAtlasColorselectedQuiescent2023}, \citet{heintzExtremeDampedLyman2023}, and \citet{brammer_2023_8319596}.
We utilized the spectroscopic redshifts from the DJA catalog to derive the rest-frame spectra for each source. 

\begin{figure*}[!htb]
    \plotone{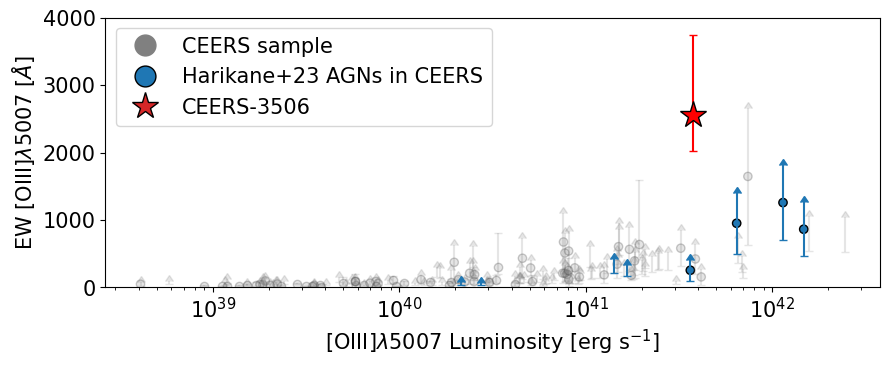}
    \plotone{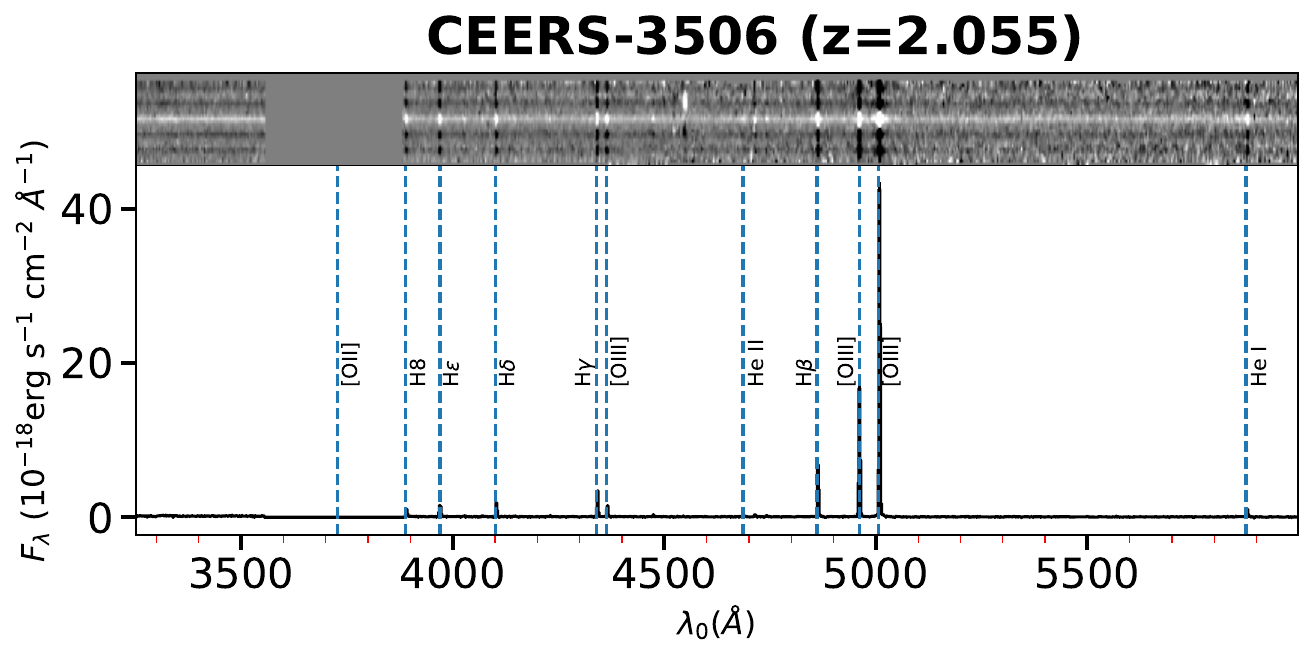}
    \caption{\textbf{Top panel}: {\OOiii} EW versus luminosity distribution of our CEERS sample. CEERS-3506 is highlighted as the red star, while other objects are in gray. The blue markers are AGNs reported in \citet{harikaneJWSTNIRSpecFirst2023}. The arrows denote 1$\sigma$ lower limits for sources whose 16th percentile of the estimated probability distribution of the continuum is lower than 0. For each object, the circle marks the median of EW value; however, if the median is less than 0 due to non-detection of the faint continuum, only the 1$\sigma$ lower limit is shown. \textbf{Bottom panel}: The 2D and 1D spectra of CEERS-3506 which include the {\Hb+{\Oiii}} emission lines, in the same manner as Figure \ref{fig:subaruspec}.}
    \label{fig:jwstspec}
\end{figure*}

We use the method for EW measurements detailed in Section \ref{sec:emission}.
Suppose the continuum is so faint that the 16th or/and 50th percentiles of the estimated probability distribution of the continuum are lower than zero. In that case, we can only establish an upper limit on the continuum, and consequently, a lower limit on the EW. Our sample is restricted to sources with at least one medium-resolution spectrum covering the wavelength range of $4800-5200$ \AA\ and the {\sc [Oiii]$\lambda$5007} line is not in the instrument gap. Out of 153 sources, one galaxy is identified with extremely high EW({\OOiii}) similar to our Subaru EELGs, exhibiting a 1$\sigma$ lower limit exceeding 1000 \AA. This finding is presented in Figure \ref{fig:jwstspec}. The basic information of the CEERS EELG is also shown in Table \ref{tab:basicinfo}.
\subsubsection{JWST spectrum}
The NIRSpec spectroscopy of this source has an ID from the micro shutter array of 3506; thus we refer to it as CEERS-3506 hereafter.  {\sc [Oiii]$\lambda$5007} falls within the range of the filter-grating pair F100LP-G140M. Figures \ref{fig:jwstspec} exhibit the 2D and 1D spectrum of this specific filter-grating pair that features the strong {\sc H$\beta$} + {\sc [Oiii]} doublet emissions. In addition, the {\sc H$\alpha$} line is captured by the F170LP-G235M spectrum. The F290LP-G395M spectrum encompasses both Pa{\sc $\beta$} and He{\sc i} 10830. However, the {\sc [Oii]$\lambda\lambda$3727,3729} doublets lie within an instrument gap of the F100LP-G140M spectrum.
\subsection{Photometric Data}
For J1000+0211, our dataset includes images from the Subaru Hyper Suprime-Cam (HSC; \citealt{aiharaHyperSuprimeCamSSP2018}) in the \textit{g}, \textit{r}, \textit{i}, \textit{z}, and \textit{y} bands, acquired as part of the third data release of HSC-SSP survey \citep{Aihara2022thirddata}. Additionally, we utilize JWST NIRcam images in the F090W, F115W, F150W, F200W, F277W, and F444W bands, and JWST MIRI images in the F770W and F1800W bands, sourced from both the COSMOS-Web (GO 1727; PI: J. Kartaltepe; \citealt{caseyCOSMOSWebOverviewJWST2023}) and PRIMER (GO 1837; PI: J. S. Dunlop; \citealt{2021jwst.prop.1837D}) programs. For CEERS-3506, our data comprises HST ACS images in the F435W, F606W, and F814W bands, HST WFC3 images in the F125W and F160W bands from the CANDLE survey \citep{groginCANDELSCOSMICASSEMBLY2011,Koekemoer_2011}, and JWST MIRI images in the F770W, F1000W, F1280W, F1800W, and F2100W bands from the CEERS survey. For J0845$-$0123, our data is limited to the Subaru HSC images.

The three sources are relatively compact and isolated. Consequently, we employ a circular aperture of approximately $1\farcs5$ in diameter for photometric analysis of all images, complemented by a slightly larger annulus for background subtraction and uncertainty estimation. The photometric measurements are performed by using the \texttt{Photutils} \citep{larry_bradley_2024_10671725}.

\section{Result}\label{sec:result}
\startlongtable
\begin{splitdeluxetable*}{lccccccccBcccccccBccccc}
\tablecaption{Selected Observed Emission-line Fluxes and EWs of Our {\sc [Oiii]} emitters\label{tab:line}}
\tablewidth{0pt}
\tablehead{
\colhead{ID} & \colhead{\sc [Oii]$\lambda\lambda$3727,29} & \colhead{\sc H$\gamma$} & \colhead{\sc [Oiii]$\lambda4363$} & \colhead{\sc H$\beta$}& \colhead{\sc [Oiii]$\lambda$4959} & \colhead{\sc [Oiii]$\lambda5007$} & \colhead{H$\alpha$} & \colhead{\sc [Nii]$\lambda$6484} & \colhead{\sc [Sii]$\lambda6717$} & \colhead{\sc [Sii]$\lambda6731$} & \colhead{\sc [Oii]$\lambda7320$} & \colhead{\sc [Oii]$\lambda7330$} & \colhead{\sc [Siii]$\lambda$9531} & Flux({\sc H$\beta$}) & \colhead{EW({\OOii})} & \colhead{EW({\sc [Oiii]$\lambda$4959})} & \colhead{EW({\OOiii})\tablenotemark{$\dagger$}} & \colhead{\sc EW(H$\beta$)} & \colhead{\sc EW(H$\alpha$)} & \colhead{EW({\sc [Oiii]$\lambda\lambda4959,5007$}+{\Hb})}\\ \colhead{} & \colhead{} & \colhead{} & \colhead{} & \colhead{} & \colhead{} & \colhead{} & \colhead{} & \colhead{} & \colhead{} & \colhead{} & \colhead{} & \colhead{} & \colhead{} & \colhead{($\rm 10^{-18}~erg~s^{-1}~cm^{-2}$)} & \colhead{(\AA)} & \colhead{(\AA)} & \colhead{(\AA)} & \colhead{(\AA)} &\colhead{(\AA)}&\colhead{(\AA)}\\
\colhead{(1)} & (2) & (3) & (4) & (5) & (6) & (7) & (8) & (9) & (10) & (11) & (12) &(13) &(14) & (15) & (16) &(17) &(18) &(19)
& (20) & (21)} 
\startdata
CEERS-3506 & -- & $45.3 \pm 1.6$ & $23.5\pm1.0$ & $100\pm2.4$ & $253\pm11$ &
$651\pm27$ & $282\pm10$ & $<$0.69 & $1.36\pm0.47$ & $2.23\pm0.56$ & $0.71\pm0.46$ & $0.42\pm0.35$ & $5.91\pm0.55$ & $22.1\pm0.5$ & -- & $984^{+574}_{-274}$ &$2508^{+1487}_{-689}$ & $341^{+133}_{-72}$ & $1618_{-378}^{+526}$ & $3868_{-1064}^{+2292}$\\
J1000+0211 & $51.0\pm7.5$ & $45.2\pm2.8$ & $16.5 \pm 2.1$ & $100\pm4.0$ & $225\pm14$ & $722\pm29$ & -- & -- & -- & -- & -- & -- & -- & $26.0\pm1.0$ & $102^{+21}_{-16}$ & $1041^{+440}_{-244}$&$2905^{+946}_{-578}$ & $246^{+51}_{-40}$ & -- & $4213_{-839}^{+1371}$\\
J0845$-$0123 & $98\pm18$ & $49.7\pm9.0$ & $12.1 \pm 2.5$ & $100\pm17$ & $232\pm5$ & $697\pm15$ & -- & -- & -- & -- & -- & -- & -- & $85.6\pm1.6$ & $210^{+31}_{-25}$ & $661^{+60}_{-51} $ &$2000^{+188}_{-159}$ & $357^{+32}_{-28}$ & -- & $2952_{-235}^{+277}$\\
\enddata
\tablecomments{The 2nd-14th columns show the line fluxes normalized by {\Hb}. ``$<$" indicates $1\sigma$ upper limit. The 16th-21st columns show the EWs. We present the median value and 16th and 84th percentiles for the EW measurements. }
\tablenotetext{\tablenotemark{$\dagger$}}{{The EWs({\OOiii}) of J1000+0211 and J0845$-$0123 are reported to be $6000\pm2000$ \AA\ and $4500\pm500$ {\AA}, respectively, in \citet{matsuokaSubaruHighExploration2018}.}}
\end{splitdeluxetable*}

\subsection{Emission Lines}\label{sec:emission}

{For {\Hb}, {\Oiii}$\lambda4959$, and {\OOiii}},
we estimate the continuum level with the range of $4800-5200$ \AA, after masking the {\sc H$\beta$} and {\sc [Oiii]$\lambda\lambda$4959,5007} doublets emission lines. 
For the three {\Oiii} emitters, in addition to masking the emission lines, we visually inspect the vicinity of {\OOiii} to identify and mask the residuals of sky subtraction or the removal of cosmic rays.
Assuming the continuum flux density remains constant locally, the continuum underlying the {\sc [Oiii]$\lambda$5007} emission line is estimated by using a Monte Carlo method to account for uncertainties in the continuum flux density. This approach involves generating 10,000 simulated datasets based on the measured continuum flux densities and their associated measurement uncertainties. The median of the posterior predictive probability distribution is adopted as the continuum level at 5007 \AA, with the 68\% confidence interval, determined by the 16th and 84th percentiles, providing the uncertainty measure. {For {\OOii}, {\Ha}, and other emission lines, the continuum level is determined separately using pixels from their respective nearby continua, excluding regions with known emission lines. The center of the selected range is adjusted slightly based on the spectral coverage limits.}

To derive the line fluxes and associated errors, 
%
%
%
%
%
we fit spectral models to the observed spectrum with the error spectrum by the package \texttt{lmfit} \citep{newville_2015_11813}. 
Here the spectral models are composed of a Gaussian model and a constant continuum:
\begin{equation}
    f(\lambda) = A \exp\left(-\frac{(x - \mu)^2}{2\sigma^2}\right) + C
\end{equation}
with 4 free parameters of amplitude $A$, line width $\sigma$, central wavelength $\mu$, and offset of continuum $C$. {The prior for the continuum offset C is set based on the values measured from the broad continuum range described in the previous paragraph.} We obtain the integrated {\OOiii} flux by integrating the 5$\sigma$ width of the Gaussian profile of the {\Oiii} line after subtracting the continuum. We compare the measured Gaussian flux with the measured integrated flux. If the two fluxes are different more than $1\sigma$ level, a second Gaussian will be added in the fitting.
We utilize the same methods to estimate the flux and EW of other lines, by masking nearby residuals and other emission lines.
Table \ref{tab:line} shows the line ratios of the major narrow component. The unresolved lines (e.g. {\OOii}) are presented with integrated measurement.
%

\subsubsection{Balmer Decrement}\label{sec:balmer}
%
To assess the effect of dust extinction, we use the Balmer line ratios of {\Hg}/{\Hb}. {\Hg} and {\Hb} are detected in all three studied objects. We obtain 
{\Hg}/{\Hb} ratios of narrow components to be $0.45\pm0.02$, $0.45\pm0.03$, and $0.50\pm0.09$, for CEERS-3506, J1000+0211, and J0845$-$0123, respectively, while the intrinsic ratio is 0.47 for $n_e = 1000~\textrm{cm}^{-3}$ and $T_e = 15,000 \textrm{K}$ \citep{osterbrock2006astrophysics}. Because the Balmer decrement ratios of {\Hg}/{\Hb} align with the intrinsic value within $1\sigma$ errors for J1000+0211 and J0845$-$0123, we do not correct for dust attenuation.
In addition, we evaluate the ratio of the narrow component of {\Ha}/{\Hb} to be $2.82\pm0.10$, where the intrinsic ratio is $2.74$ for $n_e = 1000~\textrm{cm}^{-3}$ and $T_e = 20,000 \textrm{K}$ \citep{osterbrock2006astrophysics}. Because the difference falls in the $1\sigma$ error, 
we again do not correct for dust attenuation in CEERS-3506. In summary, we consider the dust attenuation to be negligible for all three objects.

\subsubsection{Chemical Properties}
In all the objects we study, the auroral lines of {\Oiii$\lambda$4363} are detected, enabling us to calculate the oxygen abundances by using the direct method (e.g. see \citealt{isobeEMPRESSIVExtremely2022} for reference). 
Practically, we use the python package \texttt{PyNeb}\footnote{\url{https://morisset.github.io/PyNeb_Manual/html/}} \citep{luridianaPyNebNewTool2015} to conduct the calculation. 
The electron temperature of {\sc O$^{2+}$}, $T_e$({\sc Oiii}), and the electron density, $n_e$, are iteratively calculated with the emission line ratios of {\Oiii$\lambda$4363}/{\OOiii} and {\sc [Sii]$\lambda$6717}/{\sc [Sii]$\lambda$6731}, respectively. For J1000+0211 and J0845$-$0123, the {\sc [Sii]$\lambda\lambda$6717,6731} doublets are undetected. The {\OOii} doublets are unresolved in the deep VPH850 grism spectra but resolved in the VPH650 grism spectra with less exposure time (2,400 s) observed on the same nights. We adopt a standard $n_e$ of $\rm 1000~cm^{-3}$ \citep{osterbrock2006astrophysics}. The assumption leads to the ratio of {\sc [Oii]$\lambda$3727} to {\sc [Oii]$\lambda$3729} to be 1.2 at the temperature of 15,000 K\footnote{This value is obtained from \texttt{PyNeb}.}, {which is consistent with the observed line ratios in the VPH650 spectra for both objects.}
We proceed the calculation of $T_e$({\sc Oiii}) with the assumed $n_e$ and the measured line ratios of {\Oiii$\lambda$4363}/{\OOiii}. Notably, variations in $n_e$ do not largely affect the determination of $T_e$({\sc Oiii}).
We utilize the line ratios of {\OOiii}/{\Hb} to calculate the $\rm O^{2+}/H^+$ abundance at given $T_e$({\sc Oiii}) for each object.

To estimate the $\rm O^{+}/H^+$ abundance, we need to assess the electron temperature of $\rm O^{+}$, $T_e$({\sc Oii}). Because we cannot directly measure the $T_e$({\sc Oii}), we conduct an estimation by using the empirical relation of 
\begin{equation}\label{eq:te}
    T_e(\textsc{Oii}) = 0.7\times T_e(\textsc{Oiii}) + 3000
\end{equation}
\citep{garnettElectronTemperatureVariations1992}. 
For J1000+0211 and J0845$-$0123, we derive the $\rm O^{+}/H^+$ from {\OOii}/{\Hb} and $T_e$({\sc Oii}), while for CEERS-3506, we use {\sc [Oii]$\lambda\lambda$7320,30}/{\Hb} to calculate the $\rm O^{+}$ abundance.
We obtain the metallicity $\rm 12 + \log(O/H)$ by adding the abundance of $\rm O^+$ to $\rm O^{2+}$. We present the results in Table \ref{tab:properties}. 

The ionization parameter, denoted as $\log(U)$, is estimated through diagnostic line ratios. These ratios are calibrated and fitted using a bicubic surface function \citep{kewleyUnderstandingGalaxyEvolution2019}. 
\begin{align}\label{eq:logU}
    z  =& A + Bx + Cy + Dxy + Ex^2 +Fy^2+Gxy^2  \nonumber\\
     &+ Hyx^2+ Ix^3 + Jy^3
\end{align}
where the $\rm x = \log(line~ratio)$, $y=\log(O/H)+12$, and $z=\log(U)$. The coefficients ($A=13.8$, $B=9.5$, $C=-4.3$, $D=-2.4$, $E=-0.58$, $F=0.28$, $G=0.16$, $H=0.089$, $I=0.031$, and $J=0.0$) refer to \citet{kewleyUnderstandingGalaxyEvolution2019}. For J1000+0211 and J0845$-$0123, we apply the line ratio of {\OOiii}/{\OOii} (O32). For CEERS-3506, the {\OOii} lines are not covered by observation. Additionally, its metallicity falls beyond the valid calibration range established by \citet{kewleyUnderstandingGalaxyEvolution2019}. Hence the estimation of CEERS-3506 is not conducted. The chemical properties are summarized in Table \ref{tab:properties}.

\begin{table*}[!tb]
\caption{Properties of Our {\Oiii} emitters}\label{tab:properties}
\begin{ruledtabular}
    \begin{tabular}{lccc}
    & CEERS-3506 & J1000+0211 & J0845$-$0123 \\\hline
    $T_e(\textsc{Oiii}$) (K) & 20700$\pm$500 & 16000$\pm$2000 & 13700$\pm$700 \\
    $n_e$\tablenotemark{$\mathrm{\dagger}$} ($\rm cm^{-3}$) & 5000$\pm$4000  & Assumed 1000  & Assumed 1000  \\
    $\rm12 + \log(O/H)$ & 7.52$\pm$0.02 ($0.07\pm0.01Z_\odot$) & 7.81$\pm$0.12 ($0.13\pm0.04Z_\odot$) & 8.02$\pm$0.06 ($0.22\pm0.03Z_\odot$) \\
    $\log(U)$\tablenotemark{$\mathrm{\ddag}$} & -- & $-1.91$ & $-2.04$ \\
    \end{tabular}
\end{ruledtabular}
\tablenotetext{\dag}{Measured from {\sc [Sii]$\lambda\lambda$6717,6731}. For J1000+0211 and J0845$-$0123, $n_e$ is assumed to be $\rm 1000~cm^{-3}$.}
\tablenotetext{\ddag}{Measured from {\OOiii}/{\OOii} and the metallicities with \citet{kewleyUnderstandingGalaxyEvolution2019} calibration of Eq. \ref{eq:logU}.} 
\end{table*}

\subsubsection{Line Diagnostics}

\begin{figure*}[!t]
    \plotone{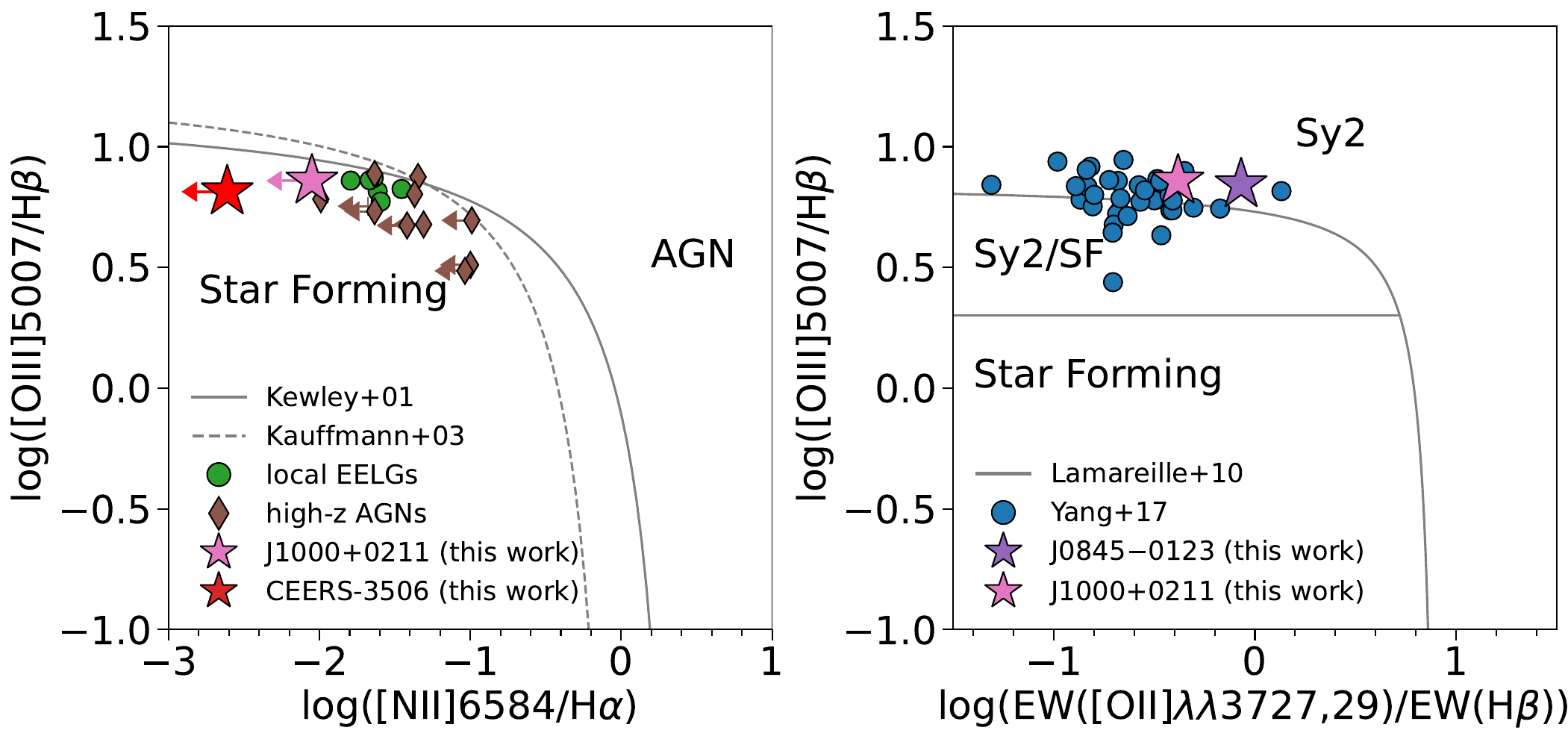}
    \caption{BPT diagram (left) and the blue diagram (right) for our objects. \textbf{The left panel}: the red and pink stars denote the results of J1000+0211 and CEERS-3506 measured in this work, respectively.
    The green circles represent local EELGs \citep{jaskotORIGINOPTICALDEPTH2013}. The brown diamonds represent high-z AGNs from \citet{harikaneJWSTNIRSpecFirst2023} and \citet{chisholmNeEmissionFaint2024}. We use the arrows if only upper limits can be constrained. The solid and dashed lines are the separation lines recommended by \citet{kewleyTheoreticalModelingStarburst2001} and \citet{kauffmannHostGalaxiesActive2003}, respectively. \textbf{The right panel}: the purple and pink stars denote the results of J0845$-$0123 and J1000+0211 measured in this work, respectively. The blue circles denote the ``blueberry" galaxies with high EW({\Oiii}) from \citet{yangBlueberryGalaxiesLowest2017}. The solid lines are the separation lines suggested in \citet{lamareilleSpectralClassificationEmissionline2010}. Similar to local EELGs and high-z AGNs, our objects are located on the border between the star formation and AGN regions in both diagrams.}
    \label{fig:diag}
\end{figure*}

Figure \ref{fig:diag} shows the BPT \citep{baldwinClassificationParametersEmissionline1981} and blue \citep{lamareilleSpectralClassificationEmissionline2010} diagram. We find the three EELGs are located near the border between star formation and AGN regions, implying that line diagnostics is insufficient to determine the physical origins.

\subsection{Broad Component Identification}
\begin{figure*}[!thb]
    \begin{center}
        \includegraphics[width=1.0\textwidth]{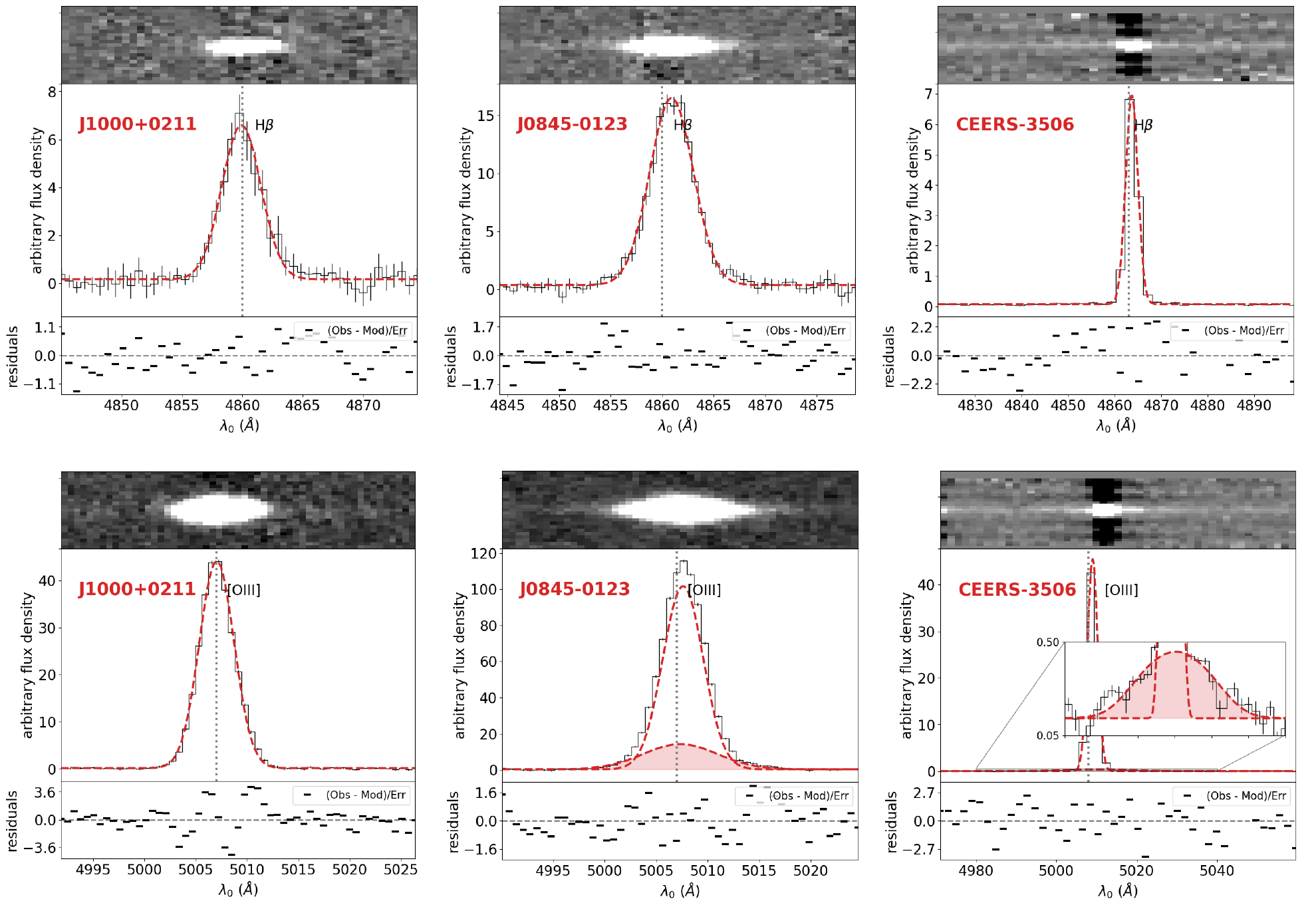}
    \end{center}
    \caption{Strong {\Hb} and {\OOiii} emission lines of the three {\Oiii} emitters. \textbf{Top panels}: {\Hb} lines for J1000+0211, J0845$-$0123, and CEERS-3506 from left to right. The black and red lines indicate observed spectra and the best-fit models, respectively. The vertical dotted lines denote the systemic redshifts. The 2D spectra and the fitting residuals are shown above and below the main panels, respectively. The horizontal dotted lines in the residual panels denote the value of 0.
    \textbf{Bottom panels}:
    %
    Same as top panels, but for {\OOiii} lines. We mark the best-fit broad components with red shades. For clarity, CEERS-3506 is shown with a detailed view in the zoom-in inset. The inset uses a logarithmic scale to highlight the faint broad component.
    The best-fit parameters for the fitting are presented in 
    Table \ref{tab:fwhm}.
    }
    \label{fig:broad}
\end{figure*}

\begin{figure*}
    \begin{center}
        \includegraphics[width=1.0\textwidth]{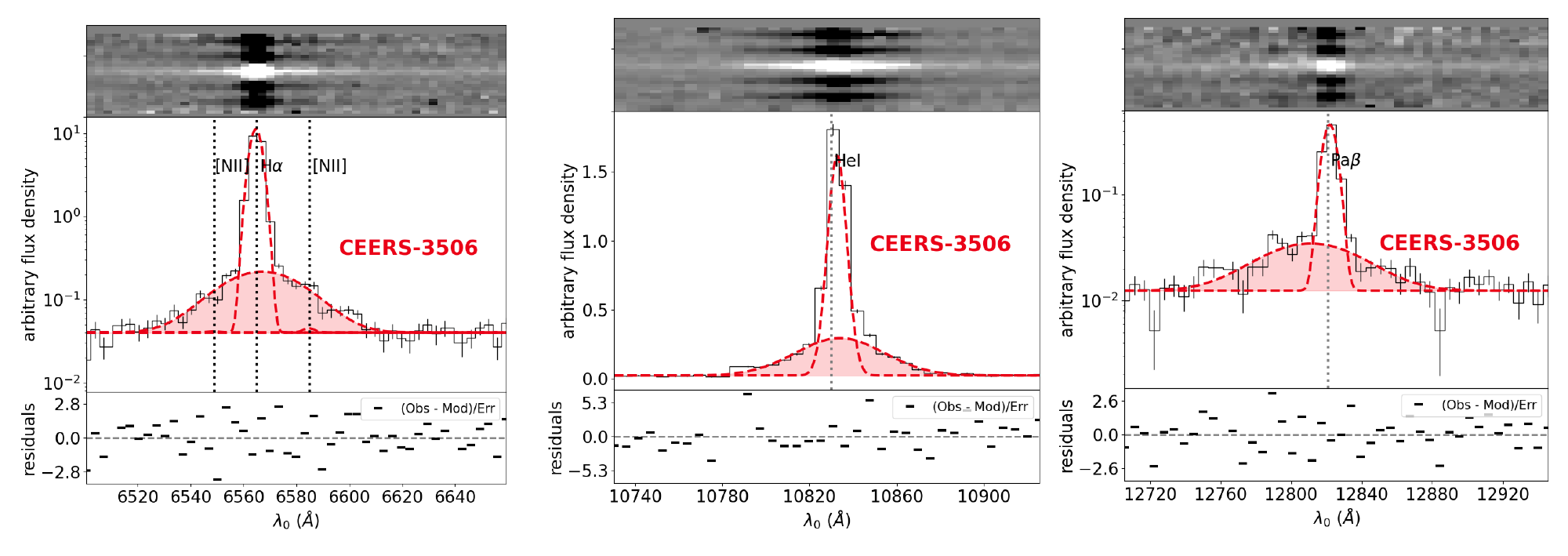}
    \end{center}
    \caption{Same as Figure \ref{fig:broad}, but for {\Ha}, He{\sc i$\lambda$10830}, and Pa{\sc $\beta$} lines of CEERS-3506 that have deep spectra that have deep spectra obtained with JWST/NIRSpec beyond the rest-frame $0.6\mu$m (cf. J1000+0211 and J0845$-$0123). The left and right panels are shown in a logarithmic scale.}\label{fig:broadceers}
\end{figure*}

In Figures \ref{fig:broad} and \ref{fig:broadceers}, we present the best-fit models and the observed spectra for the strong emission lines of our three objects.
We define a multi-Gaussian model as:
\begin{equation}
    f(\lambda) = \sum_{i=1}^N A_i \exp\left(-\frac{(x - \mu_i)^2}{2\sigma_i^2}\right) + C
\end{equation}
where $\sigma_j < \sigma_k$ if $j<k$.
In this model, \(N\) represents the number of Gaussian components, with each component defined by its amplitude \(A_i\), central wavelength \(\mu_i\), and line width \(\sigma_i\). The variable \(C\) denotes the common continuum across all components. To identify whether there exist multiple components in the emission lines, 
we use single (N=1) and double (N=2) Gaussian models, and fit the models to the strong emission lines of {\Hb} and {\OOiii} for all of the three galaxies. We additionally fit the emission lines of {\Ha}, He{\sc i$\lambda$10830}, and Pa{\sc $\beta$} for CEERS-3506, while J1000+0211 and J0845$-$0123 do not have spectra with enough sensitivity whose wavelength coverage beyond $\sim$ 5500 {\AA} in the rest frame. For simplicity, we refer to the 1st and 2nd Gaussian components in the double Gaussian model as the narrow and broad components, respectively. In addition, we require that the full width at half maximum (FWHM), determined by \({\rm FWHM}=2\sqrt{2\ln{2}} \cdot \sigma\) for the Gaussian distribution, satisfies \(\rm FWHM_{narrow}<500~km~s^{-1}\) for the narrow component and \(\rm FWHM_{broad}>500~km~s^{-1}\) for the broad component.

%
%
%
To choose the best models from the single and double Gaussian models, we use
Akaike Information Criterion (AIC; \citealt{akaikeNewLookStatistical1974}) that is defined by $ \text{AIC} = -2\log(L) + 2k$.
Here, $L$ represents the likelihood of the model, and $k$ denotes the number of parameters in the model. 
We define the criteria for an emission line to be better explained by a double Gaussian model with $\rm \Delta AIC_{double - single} < -10$ and $\rm S/N_{broad}>5$, where $\rm \Delta AIC_{double - single}$ is defined by the AIC of the double Gaussian model subtracted by that of the single Gaussian model and $\rm S/N_{broad}$ is the signal-to-noise ratio of the broad component. 
We confirm that these criteria work properly by visually inspecting the distribution of the residuals of data to the best-fit double Gaussian model.
%
%
%
%


%
%

%


%
The broad component is attributed to the broad-line region (BLR) of AGN or the galactic outflow because the gas motion of the BLR/outflow is faster than the interstellar medium (ISM) and/or narrow-line region (NLR) which are represented by the narrow component. 
Note that a type 1 AGN does not show broad forbidden lines (e.g. {\OOiii}) but broad permitted lines (e.g. {\Hb} and {\Ha}),
because the electron density in the BLR is higher than the critical densities of forbidden lines.
Unlike BLRs, outflows produce both broad forbidden lines and broad permitted lines because the electron density is lower than the critical density. 
%
%
%
%

%
\subsubsection{{\OOiii} Lines}
Because the BLR of AGN does not produce the broad line feature in {\Oiii}, we search for outflow signatures with the strong {\OOiii} emission lines in our three galaxies. We evaluate the $\rm \Delta AIC_{double-single}$ values to be $+5.3$, $-75.8$, and $-68.1$ for J1000+0211, J0845$-$0123, and CEERS-3506, respectively, while our criterion is $\rm \Delta AIC_{double-single}<-10$. We obtain the $\rm S/N_{broad}$ to be 5.7 for J0845$-$0123 and 12.6 for CEERS-3506, which satisfy our requirement for the $\rm S/N$ ($\rm S/N_{broad}>5$). Therefore, we conclude that one out of the three EELGs, J1000+0211,
does not have a broad component in {\OOiii} emission lines. The other two galaxies, J0845$-$0123 and CEERS-3506, have
the broad components with
$\rm FWHM_{outflow} = FWHM_{\rm broad,\text{\Oiii}}=507\pm38~km~s^{-1}$ and $\rm 1093\pm107~km~s^{-1}$, respectively. The details of fitting results are presented in Table \ref{tab:fwhm}. 
{Moreover, we fit the {\Oiii}$\lambda$4959 with the same FWHM as that of {\OOiii}. We find the flux ratios of the broad {\Oiii}$\lambda$4959 to {\OOiii} lines are $3.1\pm1.1$ and $3.2\pm1.4$ for J0845$-$0123 and CEERS-3506, respectively. These values are consistent with the intrinsic ratio of 3 predicted by atomic physics, further confirming the presence of a broad component in the [O III] lines for both objects.}
Notably, the outflow velocity of CEERS-3506 is larger than a typical starburst-driven outflow velocity ($\rm \sim500 ~km~s^{-1}$; \citealt{heckmanSystematicPropertiesWarm2015,veilleuxGalacticWinds2005} ), suggesting the presence of other drivers (e.g. AGN) in CEERS-3506.
\subsubsection{{\Hb} Lines}
For the one galaxy with no outflow signature J1000+0211, we apply the single and double Gaussian models to the {\Hb} line. The results prefer the single Gaussian model ($\rm \Delta AIC_{double - single} = 5.3$).
%
%
For the two galaxies with outflow signatures, J0845$-$0123 and CEERS-3506, we calculate the $\rm \Delta AIC_{double - single}$ values for the {\Hb} lines to be $-12.4$ and $-6.5$, respectively. Notably, CEERS-3506 does not satisfy our AIC criterion of $\rm \Delta AIC_{double - single} < -10$. The $\rm S/N_{broad}$ values for the {\Hb} lines are 2.2 and 3.2, both of which fall below the significance threshold of $\rm S/N > 5$. Thus, we conclude that no reliable broad components in {\Hb} are detected for either J0845$-$0123 or CEERS-3506.
We summarize the parameters of the best-fit models in Table \ref{tab:fwhm}.

%
%
%
%
%
\subsubsection{Other Lines of CEERS-3506}\label{sec:broadother}
For CEERS-3506, the spectra are sensitive enough for us to investigate other permitted lines, {\Ha}, He{\sc i$\lambda$10830}, and Pa{\sc $\beta$}. We search for the outflow and/or BLR emission of these permitted lines in CEERS-3506.
%
%

We have detected a broad component in {\OOiii}, suggesting the presence of an outflow in CEERS-3506. 
However, the broad-to-narrow flux ratio of {\OOiii} is notably low at 0.05. There is a possibility that the outflow signals may not be detected in other lines (like {\Hb} discussed above; see \citealt{carnianiIonisedOutflowsQuasar2015}). Hence, out of practical feasibility, we first compare the single and double Gaussian models. If the double Gaussian model fits better than the single Gaussian model, we then conduct the comparative analysis with the double and triple (N=3) Gaussian models. Here we define $\rm \Delta AIC_{triple - double}=AIC_{triple} - AIC_{double}$, and require $\rm \Delta AIC_{triple - double} < 0$ to choose the triple Gaussian model as the best fit.

%
%
%
%
%

For {\Ha} of CEERS-3506, we simultaneously fit {\Ha} and {\sc [Nii]$\lambda\lambda$6548,84} lines. 
We fix the wavelength difference between {\sc [Nii]$\lambda\lambda$6548,84} and {\Ha}, set the FWHMs of {\sc [Nii]$\lambda\lambda$6548,6584} to be the same as the narrow component of H$\alpha$, and fix the flux ratio of {\sc [Nii]$\lambda$6548} to {\sc [Nii]$\lambda$6584} at 0.327 \citep{osterbrock2006astrophysics}, as this ratio is insensitive to both electron temperature and density.
%
%
For all of the permitted lines of {\Ha}, He{\sc i$\lambda$10830}, and Pa{\sc $\beta$} of CEERS-3506, we calculate the $\rm \Delta AIC_{double - single} = AIC_{double} - AIC_{single}$ values to be $-85.1$, $-77.9$, and $-40.7$, and find that all of these lines have significant broad components with $\rm S/N_{broad}=13.6$, 18.7, and 7.1, respectively.
%
%
%
%
All of these three lines meet our criteria for preferring the double Gaussian models ($\rm \Delta AIC_{double - single}<-10$ and $\rm S/N_{broad}>5$; see above). We proceed with the analysis of the triple Gaussian models. Consequently, we obtain the values of $\rm \Delta AIC_{triple - double}$ to be $+2.4$, $+3.8$, and $+6.5$, for {\Ha}, He{\sc i$\lambda$10830}, and Pa{\sc $\beta$}, respectively.
The values of $\rm \Delta AIC_{triple - double}$ do not meet the criterion of $\rm \Delta AIC_{triple - double} < 0$, we conclude that the triple Gaussian models are not preferable for the three permitted lines, but the double Gaussian models.
We show the fitting results in Table \ref{tab:fwhm}.

The $\rm FWHMs_{broad}$ are $1659\pm116$, $1325\pm67$, and $1513\pm219$ $\rm km~s^{-1}$ for {\Ha}, He{\sc i$\lambda$10830}, and Pa{\sc $\beta$}, respectively, 
while $\rm FWHM_{\rm outflow}$ is $\rm 1093\pm107~km~s^{-1}$ in the {\OOiii} line. 
%
Additionally, the broad-to-narrow flux ratios of {\Ha} and Pa{\sc $\beta$} are 0.11 and 0.35, respectively, while that of {\OOiii} is 0.05. Typically, outflows show a stronger broad-to-narrow ratio for {\OOiii} compared to {\Ha} \citep{marshall2023ga}.
The differences in FWHMs between the {\OOiii} line and the Hydrogen lines ({\Ha} and Pa{\sc $\beta$}), as well as the smaller broad-to-narrow ratios for the {\OOiii} line compared to the Hydrogen lines, suggest the different origins of the broad components detected in the Hydrogen lines from that of {\OOiii}. The He{\sc i$\lambda$10830} line is influenced by multiple factors, including recombination, collisional excitation, and its optical thickness, complicating the interpretation of its broad-to-narrow ratio. However, the detection of a high-velocity broad component in He{\sc i$\lambda$10830} is consistent with the presence of an AGN.

%
In addition, we find that the broad-to-narrow ratios of the Hydrogen lines increase as the wavelength increases ({\Ha}: $0.11\pm0.01$; Pa$\beta$: $0.35\pm0.05$), implying that the dust attenuation of broad lines differs from that of narrow lines. This phenomenon can be better explained by the scenario in which the broad components of the Hydrogen lines come from the BLR partly obscured by the dust torus, rather than attributing the broad Hydrogen lines to outflow emission.

In summary of our analysis of the three permitted lines, we decide to choose the double Gaussian model as the best-fit model for {\Ha}, He{\sc i$\lambda$10830}, and Pa{\sc $\beta$} of CEERS-3506, and we prefer the hypothesis that the broad components of the Hydrogen lines come from the BLR. However, whether the origin of the broad components in the Hydrogen lines is the outflow or BLR does not change our main conclusion that CEERS-3506 harbors AGN.

\begin{table*}[!bh]
\caption{Best-fit Line Profiles of Our {\Oiii} Emitters\label{tab:fwhm}}
\begin{ruledtabular}
\begin{tabular}{lcccccccc}
  \colhead{ID} & \colhead{Line} & \colhead{$\rm FWHM_{narrow}$} & \colhead{$\rm FWHM_{broad}$} & \colhead{$\rm S/N_{broad}$} &\colhead{$\rm Flux_{broad}/Flux_{narrow}$} & \colhead{$\Delta v_{\rm broad}$} & \colhead{$\rm \Delta AIC$}\\
   & & [$\rm km~s^{-1}$]&[$\rm km~s^{-1}$]& & & [$\rm km ~s^{-1}$]&\\
  \colhead{(1)} & \colhead{(2)} & \colhead{(3)} & \colhead{(4)} & \colhead{(5)} & \colhead{(6)} & \colhead{(7)} & \colhead{(8)}\\ \hline
  J1000+0211 & {\Hb} & $235\pm9$ & -- & -- & -- & -- & +5.3\\
  {} & {\OOiii}& $240\pm2$ & -- & -- & -- & -- & +1.2\\\hline
  J0845$-$0123 & {\Hb} & $272 \pm 17$ & -- & -- & -- & -- & $-12.4$\\
  {} & {\OOiii} & $273\pm6$ & $507\pm38$ & 5.7 & $0.25\pm0.05$ & $-16\pm7$ &$-75.8$\\\hline
  CEERS-3506 & {\Hb} & $228\pm4$ & -- &--& -- & -- &$-6.5$\\
  {} & {\OOiii} & $214\pm4$ & $1093\pm107$ & 12.6 & $0.04\pm0.00$ & $74\pm34$ &$-68.1$\\
  {} & {\Ha} & $241\pm3$ & $1659\pm 116$ & 13.6  & $0.11\pm0.01$ & $108\pm40$ &$-85.1$\\
  {} & He{\sc i$\lambda$10830} & $296 \pm13$ & $1325\pm67$& 18.7 & $0.77\pm0.05$ & $-10\pm50$ & $-77.9$\\
  {} & Pa{\sc $\beta$} & $214\pm7$ & $1513\pm219$ & 7.1 & $0.35\pm0.05$ & $-260\pm110$ & $-40.7$
\end{tabular}
\end{ruledtabular}
\tablecomments{(3) FWHM of the narrow component. (4) FWHM of the broad component ($\rm FWHM_{broad}>FWHM_{narrow}$). (5) S/N of the broad component. (6) The flux ratio between the broad and narrow components. {(7) $(\mu_{\rm broad} - \mu_{\rm narrow})/\mu_{\rm narrow} \times c$: the central velocity offset of the broad component compared to the narrow component.} (8) $\rm AIC_{double} - AIC_{single}$: a more negative value indicates a stronger preference for the double Gaussian model over the single Gaussian model.}
\end{table*}

\subsection{SED Fitting and Photometric Result}\label{sec:sed}
\begin{figure*}[!hb]
    \plotone{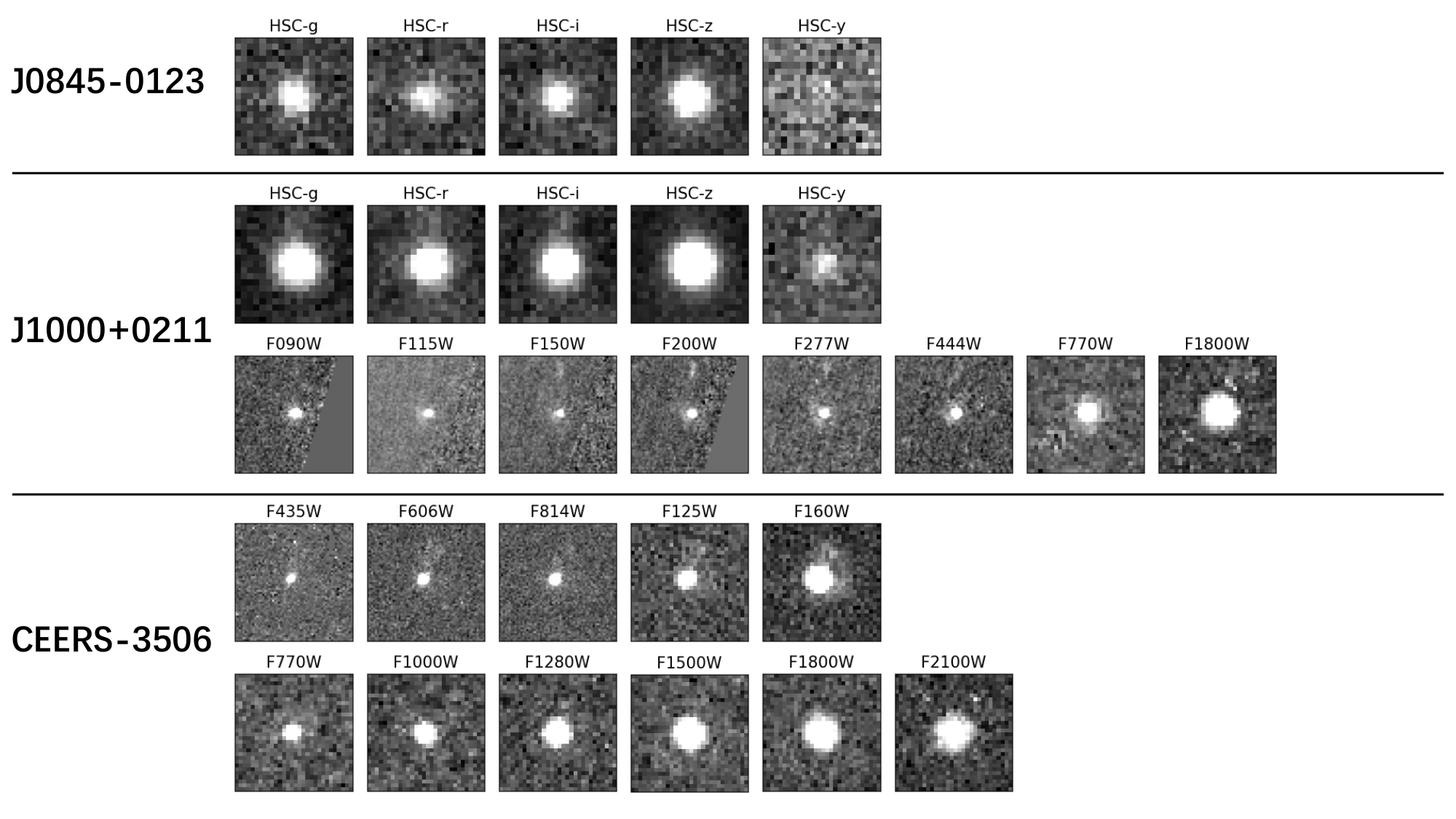}
    \caption{$1\farcs5 \times 1\farcs5$ images of our {\sc [Oiii]} emitters. The Subaru/HSC \textit{g, r, i, z, y} thumbnail cutouts are shown for J0845$-$0123. We collect the Subaru/HSC \textit{g, r, i, z, y} - bands, the JWST/NIRCam F090W, F115W, F150W, F200W, F277W, F444W, and the JWST/MIRI F770W, F1800W thumbnail cutouts for J1000+0211. For CEERS-3506, the JWST/NIRCam images are not available, so we instead show the HST/ACS F435W, F606W, F814W, and the HST/WFC3 F125W, F160W images together with the JWST/MIRI F770W, F1000W, F1280W, F1500W, F1800W, and the F2100W cutouts.}\label{fig:thumb}
\end{figure*}

\begin{figure*}[!tb]
    \centering
    \includegraphics[width=1\textwidth]{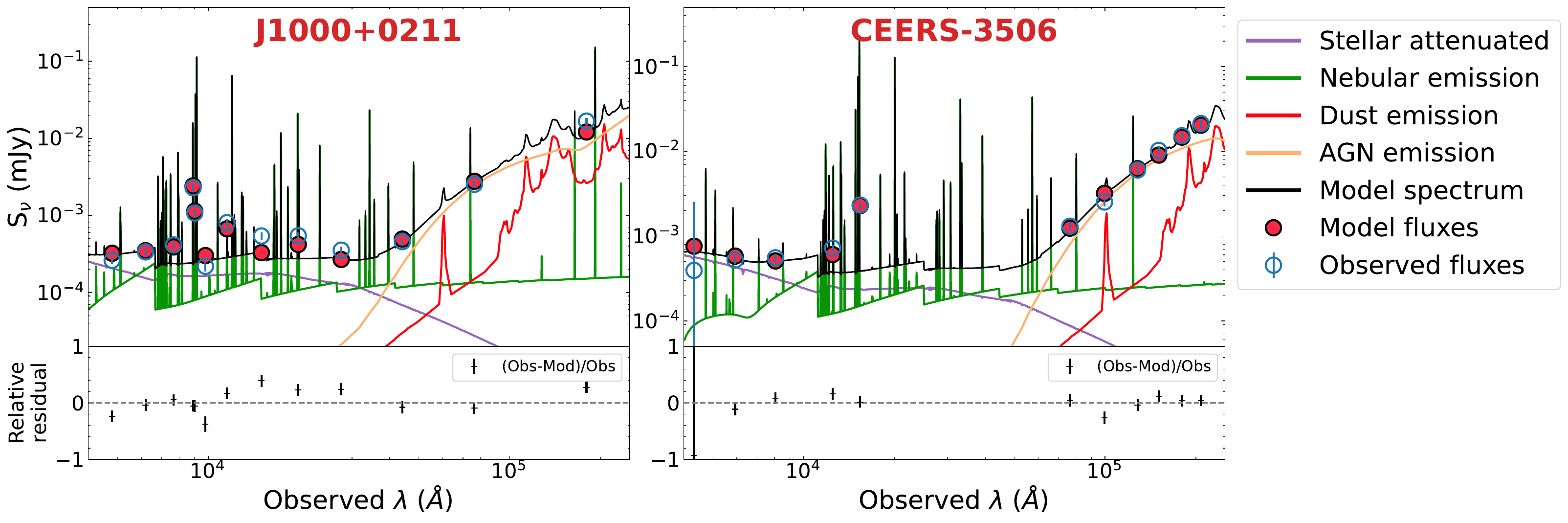}
    \caption{\texttt{CIGALE} SED fitting results of the two {\sc [Oiii]} emitters, J1000+0211 (left) and CEERS-3506 (right). The blue open and red filled circles mark the observed fluxes and the model fluxes, respectively. The solid lines mark the best-fit models of the total spectrum (black), stellar continuum (purple), nebular emission (green), AGN emission (orange; including emission from surrounding dust torus), and the dust emission (red; dust heated by stars).
    Both J1000+0211 and CEERS-3506 display the near-infrared excess ($\gtrsim$ 2 $\mu$m in the rest-frame) suggesting the existence of hidden AGN complemented by our spectroscopic measurements. The bottom panels are the relative residuals.}\label{fig:SED}
\end{figure*}

\begin{figure}[!tb]
    \plotone{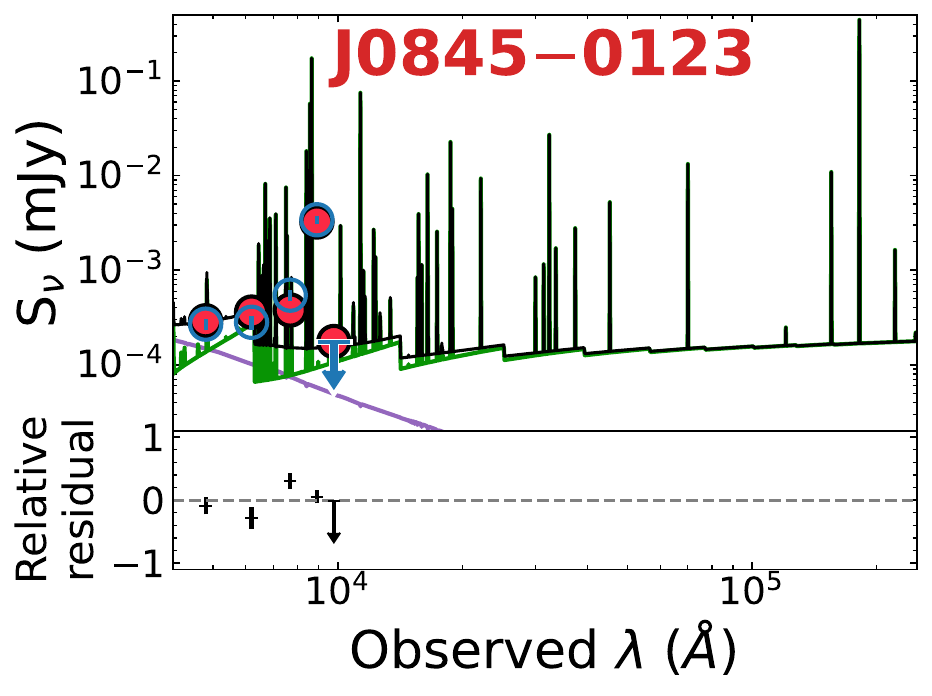}
    \caption{Same as Figure \ref{fig:SED} but for J0845$-$0123. The arrow indicates a 1$\sigma$ upper limit.}
    \label{fig:seda46}
\end{figure}

Figure \ref{fig:thumb} shows the cutouts of our three {\sc [Oiii]} emitters in multi-wavelength bands. 
We use the \texttt{Photutils}\footnote{\url{https://photutils.readthedocs.io/}} python package to perform the aperture photometry measurement and error estimation.
%
%
We employ the \texttt{CIGALE} code \citep{boquienCIGALEPythonCode2019} to conduct the SED fitting. The \texttt{CIGALE} code obtains the best-fit model spectra with the least reduced $\chi^{2}$ method. For J1000+0211 and CEERS-3506, we have multi-wavelength photometry data from rest-frame UV to mid-infrared. To exploit this rich dataset, we combine the following \texttt{CIGALE} modules:  \texttt{sfhdelayed}, \texttt{bc03}, \texttt{nebular}, \texttt{dustatt\_modified\_starburst}, \texttt{dale2014}, \texttt{skirtor2016}, \texttt{restframe\_parameters}, and \texttt{redshifting}. With these modules, our models combine stellar emission, nebular emission, dust emission, AGN contribution, and dust attenuation. We perform SED fitting to complement our spectroscopic results, primarily to constrain stellar properties and secondarily to detect AGN IR emission. {To confirm that the IR excess originates from the AGN, we include dust emission even though the dust content is suggested to be negligible in our spectroscopic result.} The parameters for redshift, metallicity, and dust attenuation are fixed based on spectroscopic measurements. The specific parameter configurations we focus on are detailed in Table \ref{tab:modules}. We adopt the default values from \texttt{CIGALE} for other parameters.
\begin{table*}
\begin{deluxetable*}{lll}
\tablecaption{Initial setup for \texttt{CIGALE} modules\label{tab:modules}}
\tablehead{
Parameter & Description & Values
}
\startdata
tau\_main & e-folding time (\(\tau\)) of the main stellar population model in Myr & 50, 100, 200, 500, 1000,\\&& 2000, 4000, 8000, 10000,\\&& 20000, 50000 \\ 
age\_main & Age of the main stellar population in the galaxy in Myr & 1, 2, 3, 4, 5, 6, 7, 8, 9, 10 \\ 
imf & Initial mass function & \citet{Chabrier_2003} \\ 
f\_burst & Mass fraction of the late burst population & 0.0 \\ 
fracAGN & Fraction of AGN IR luminosity to total IR luminosity & 0.1, 0.2, 0.3, 0.4, 0.5, 0.6,\\&& 0.7, 0.8, 0.9, 0.99 \\ 
oa & Angle measured between the equatorial plane and edge of the torus & 40 \\ 
i & Viewing angle, the position of the instrument with respect to the AGN axis & 90, 70, 50 \\
alpha & alpha slope of dust emission in the \citet{Dale_2014} model. & 1.0, 2.0, 4.0
\enddata
\end{deluxetable*}
\tablecomments{We use the module \texttt{sfhdelayed} and ${\rm SFR}(t) \propto \frac{t}{\tau^2} \exp(-t/\tau)$}
\end{table*}

We present model spectra in Figure \ref{fig:SED} and list the properties derived from the SED analysis in Table \ref{tab:sed}. We use the stellar age constrained by SED fitting in Figure \ref{fig:cloudysf}.
Most importantly, we detect the near-infrared excess in the MIRI data for both objects, while we infer the negligible dust content from the Balmer decrements of the narrow line components (see Section \ref{sec:balmer}). This indicates the existence of AGNs for both objects, which supports our argument in Section \ref{sec:broadother}. Plus, our SED fitting results suggest the two systems are AGN-dominated with fractions of AGN infrared luminosity to total infrared luminosity at 0.6 and 0.8 for CEERS-3506 and J1000+0211, respectively.
Assuming the Eddington ratios to be 1, we utilize the bolometric luminosity of the SED models to calculate the black hole masses of CEERS-3506 and J1000+0211. The resulting black hole masses are $\log (M_{\rm \bullet, SED}/M_\odot) = 6.70$ and 5.95, respectively.

For J0845$-$0123, due to the absence of the photometric data beyond the rest-frame 0.6$\mu$m that are key to constraining the contribution of long-lived, low-mass stellar populations, a precise estimate on the physical properties with SED fitting is challenging. 
We present the model spectra in Figure \ref{fig:seda46} as a reference. 
It is crucial to note that the properties of J0845$-$0123 carry significant systematics due to the inadequacy of photometric data points.

\begin{table*}[htb!]

    \caption{Properties of Our {\Oiii} Emitters Derived from SED Fitting}
    \begin{ruledtabular}
    \begin{tabular}{l c c c c c c}
    ID & SFR & Age & $\log(M_*/M_\odot)$ & $\log ({M_{\rm \bullet, SED}}/{M_\odot})$\tablenotemark{b} & $f_{\rm AGN}$\tablenotemark{c} & reduced $\chi^2$ \tablenotemark{d}\\
      & $\rm (M_\odot~yr^{-1})$& (Myr) &  &  & \\\hline
    CEERS-3506 & $60\pm35$& $2.0\pm0.7$ & $8.07\pm0.03$ & $6.70\pm0.02$ & 0.6 &1.2 \\
    J1000+0211 & $3.0\pm0.3$ & $3.1\pm0.3$ & $7.08\pm0.02$ & $5.95\pm0.07$ & 0.8 & 4.1\\
    J0845$-$0123\tablenotemark{a} & $6.1\pm2.7$& $1.5\pm0.5$ & $6.96\pm0.06$ & -- & -- & 2.7 \\
    \end{tabular}
    \end{ruledtabular}
    \label{tab:sed}
    \tablenotetext{a}{The derived properties of J0845$-$0123 have significant systematics because of the lack of the rest-frame infrared bands.}
    \tablenotetext{b}{This value is calculated based on the total disk luminosity estimates from the AGN emission module (\texttt{skitor2016}) of \texttt{CIGALE}. We assume an Eddington ratio of 1 to calculate the referential values.}
    \tablenotetext{c}{Fractions of the AGN infrared luminosity to the total infrared luminosity.}
    \tablenotetext{d}{Reduced $\chi^2$ values for the best-fit SED models.}
\end{table*}
\subsection{Surface Brightness Profiles}\label{sec:sb}
\begin{figure*}
    \plotone{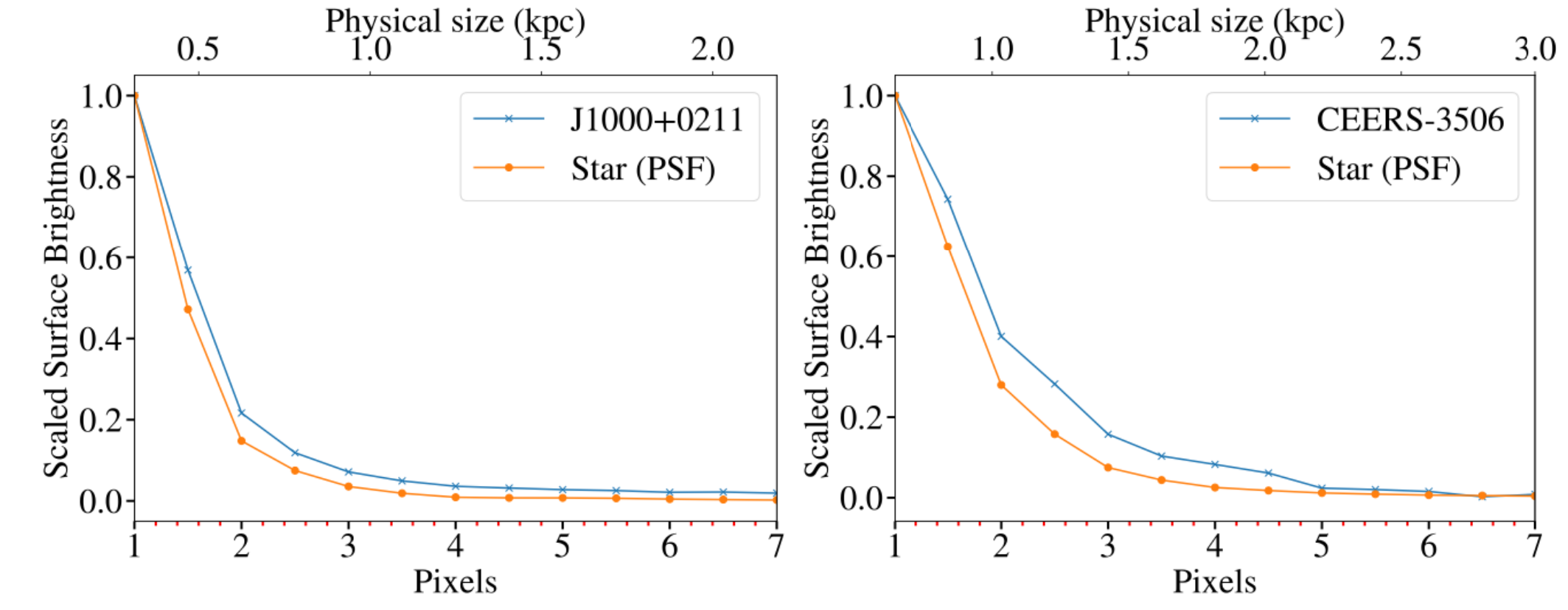}
    \caption{Surface brightness profiles (annuli) of CEERS-3506 (left) and J1000+0211 (right) compared with nearest stars. The blue crosses mark the surface brightness within the evenly spaced annuli. The orange crosses denote the closest star representing PSF. The comparison is conducted with the ACS-F435W ($\lambda_0: 0.13-0.16\mu$m) image and NIRCam-F150W ($\lambda_0: 0.73-0.91\mu$m) image for CEERS-3506 and J1000+0211 respectively. The filters are chosen due to less contamination from the emission lines. This comparison reveals the compactness of J1000+0211 and CEERS-3506.}\label{fig:psf}
\end{figure*}

We compare the objects' surface brightness (SB) profiles with the point-spread function (PSF). 
This analysis is limited to space telescope images to enhance spatial resolution. 
To minimize the emission line contamination from the {\Oiii} and Balmer lines, we use the HST ACS-F435W image (rest frame: $0.13-0.16$ $\mu$m) for CEERS-3506 and the JWST NIRCam-F150W image (rest frame: $0.73-0.91$ $\mu$m) for J1000+0211, respectively. We do not perform the analysis for J0845$-$0123 because we only have ground-based Subaru-HSC photometric data for it. We employ the \texttt{Galight} package for SB measurement  (\citealt{dingGalaxyShapesLight2021}, \citealt{dingOpeningEraQuasarhost2022}).
To obtain the empirical PSF of the two objects, we select the nearest stars with the \texttt{Galight} search algorithm\footnote{\url{https://github.com/dartoon/galight}} and subsequent manual inspection. For J1000+0211, we verify that our PSF candidate is a star in the COSMOS2020 catalog \citep{weaverCOSMOS2020PanchromaticView2022}. The radial profile comparisons between our {\Oiii} emitters and these PSF references are illustrated in Figure \ref{fig:psf}. Plus, we find that changing the PSF representative does not make a large difference in our comparison.

Based on the surface brightness profiles obtained in Figure \ref{fig:psf}, we calculate the concentration index defined as $c=R_{90}/R_{50}$ \citep{Strateva_2001_color,Shimasaku_2001_stast,shenSizeDistributionGalaxies2003}, where the $R_{90}$ and $R_{50}$ are the radii enclosing 90 and 50 percent of the total flux, respectively. The concentration indices are estimated to be 2.7 and 3.7 for CEERS-3506 and J1000+0211, respectively, while an exponential disc has $c\sim 2.3$ \citep{shenSizeDistributionGalaxies2003}. Thus we conclude that both objects exhibit compact morphology. The compactness is consistent with the high AGN fraction (CEERS-3506: 0.6; J1000+0211: 0.8; see Section \ref{sec:sed}) suggested by our SED fitting results. 

\section{Discussion}\label{sec:discuss}
%
%
%
%

\subsection{AGN Properties}
\label{sec:AGNceers}

%
%

\begin{figure}[!tb]
    \plotone{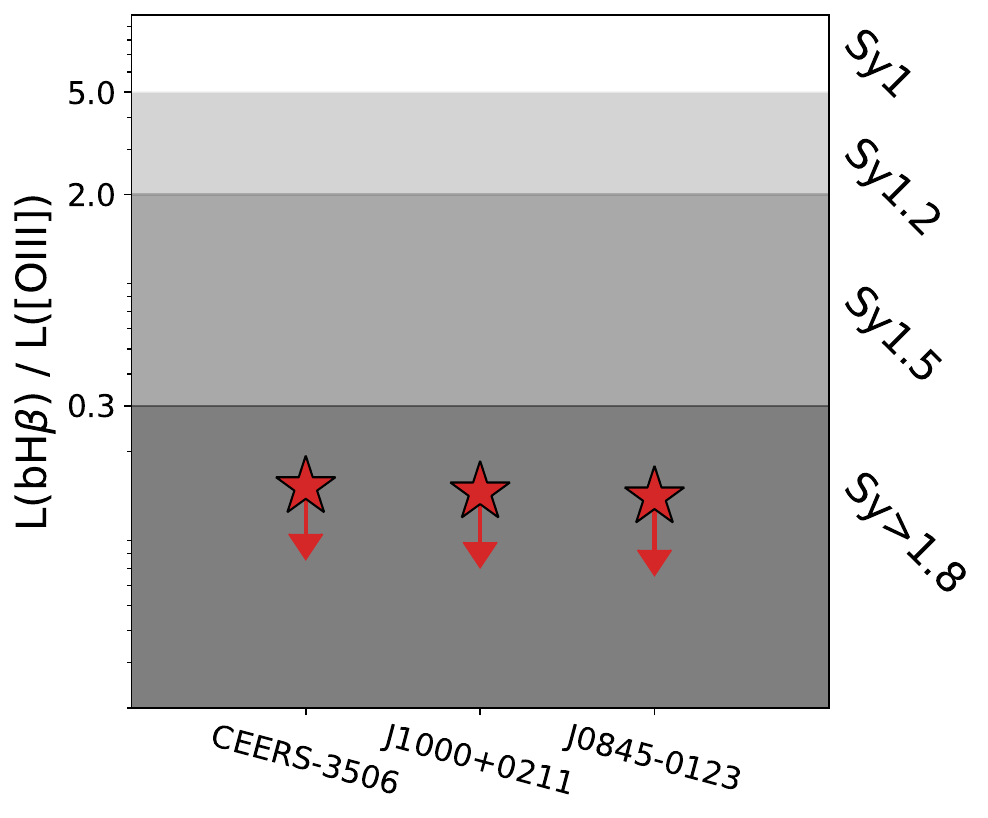}
    \caption{Quantitative classification of AGN subtypes based on the luminosity ratio between the broad {\sc H$\beta$} (b{\Hb}) emission and the {narrow} {\sc [Oiii]$\lambda$5007} emission (\citealt{winklerVariabilityStudiesSeyfert1992}; \citealt{mejia-restrepoBASSXXVDR22022}). On the y-axis, we represent this luminosity ratio with the logarithmic scale. The x-axis marks the studied objects' IDs. Since we do not detect broad {\Hb} emission in our objects, we plot upper limits using the ratio of the total {\Hb} emission (including the narrow line) to the {\OOiii} emission line.}
    \label{fig:Syclass}
\end{figure}

Figure \ref{fig:Syclass} presents the line ratios of broad {\Hb} (b{\Hb}) and {\OOiii}. The line ratios are related to the AGN subtypes \citep{winklerVariabilityStudiesSeyfert1992}. The line ratios of Sy1, Sy1.2, Sy1.5, and Sy$>$1.8 are indicated in Figure \ref{fig:Syclass}, where Sy$>$1.8 corresponds to Sy1.8, Sy1.9, and Sy2.
We calculate the ratio of the total integrated {\Hb} flux to {\OOiii} flux to be 0.15, 0.14, and 0.14 for CEERS-3506, J1000+0211, and J0845$-$0123, respectively, which are used as the upper limit of the ratio of broad {\Hb} luminosity to {\OOiii} luminosity.
Comparing the line ratios for these subtypes with our three objects, we find that all of our three objects are classified as Sy$>$1.8 subtype 
in case that all of the three objects are AGN.


%
%


We then discuss AGN properties of CEERS-3506 having the broad {\Ha} and Pa{\sc $\beta$} lines that allow us to estimate black hole masses.
%
In the following analysis, 
we assume that the broad components in the {\Ha} and Pa{\sc $\beta$} lines originate from the BLR (Section \ref{sec:broadother}).

\begin{figure*}[!tbh]
    \plotone{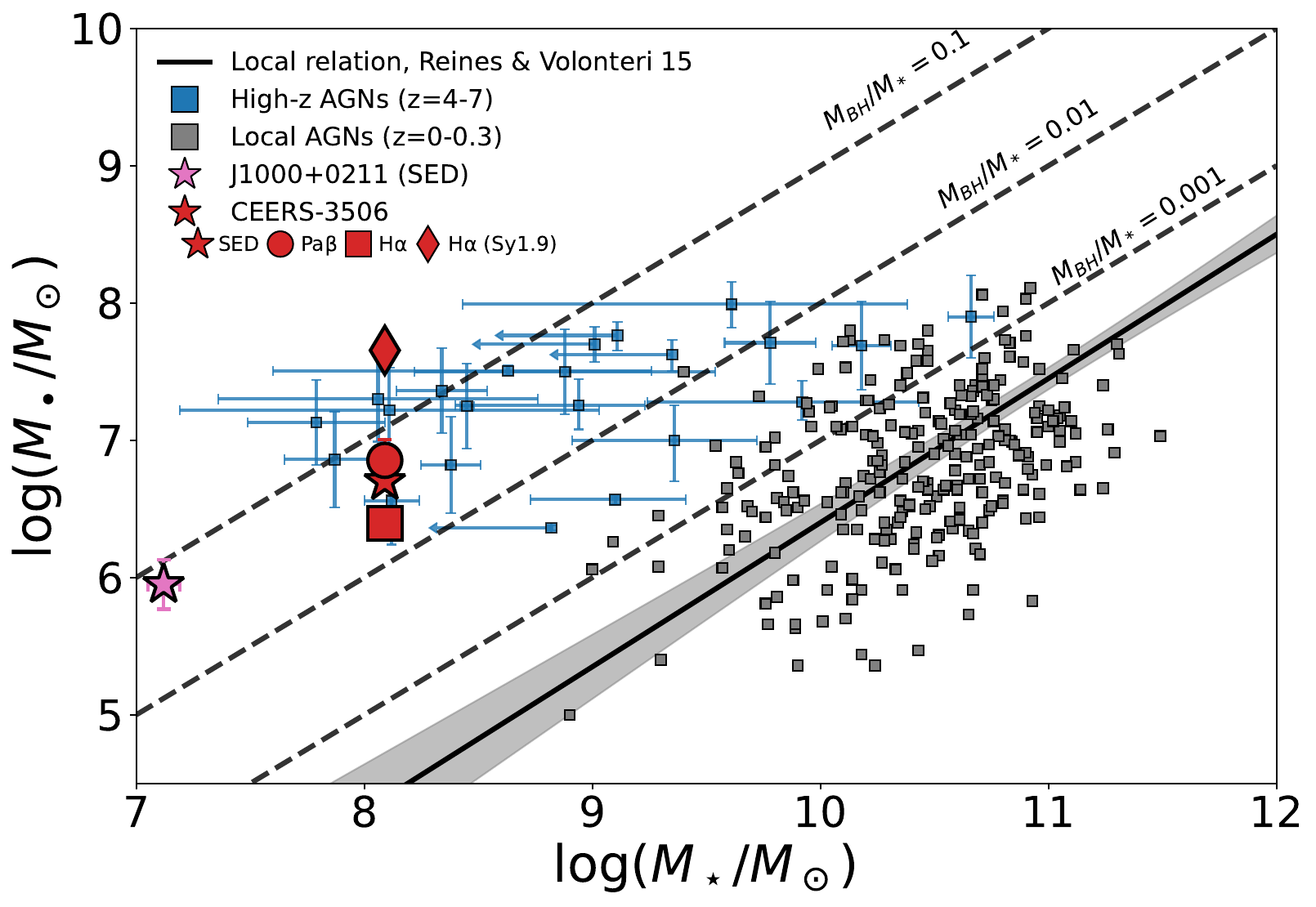}
    \caption{Relation between black hole mass ($M_\bullet$) and host's stellar mass ($M_*$). The red symbols mark the black hole mass of CEERS-3506 estimated by various methods (star: SED, by assuming $\lambda_{\rm Edd}=1$; circle: Pa{\sc $\beta$} calibration; square: {\Ha} calibration; diamond: {\Ha} calibration with Sy1.9 correction). The pink star marks the estimation of J1000+0211 (SED, the same method as CEERS-3506). The blue squares are AGNs at $z=4-7$ from \citet{harikaneJWSTNIRSpecFirst2023} and \citet{maiolinoJADESDiversePopulation2023}. The gray squares mark the local AGNs from \citet{chenHardXraySelected2017} and \citet{reinesRelationsCentralBlack2015}. The black line is the local relation \citep{reinesRelationsCentralBlack2015} with 1$\sigma$ region marked by gray shade. The dashed lines denote the $M_\bullet/M_*=$ 0.001, 0.01, and 0.1.
    This suggests that our studied objects are AGN-dominated and show more similarity to the AGNs at high redshift.}\label{fig:mass}
\end{figure*}

We present the black hole mass of CEERS-3506 estimated by various methods in Table \ref{tab:BH} and display the black hole masses and stellar masses in Figure \ref{fig:mass}.
We utilize the relation calibrated at $z\sim0$ in \citet{greeneEstimatingBlackHole2005} for the estimation of the black hole mass,
\begin{align}
    M_\bullet &= 2.0^{+0.4}_{-0.3}\times10^6M_\odot \nonumber\\ 
    &\times \left (\frac{L_{\text{{\Ha},broad}}}{10^{42}{\rm{erg~s^{-1}}}}\right )^{0.55\pm0.02} \left({\rm \frac{FWHM_{\text{{\Ha},broad}}}{\rm 10^3km~s^{-1}}}\right)^{2.06\pm0.06}.
\end{align}
We estimate the value of $\log ({M_\bullet}/{M_\odot})$ to be $6.4\pm0.1$. For comparison, we show the result applying a correction for Sy1.9 AGNs suggested in \citet{mejia-restrepoBASSXXVDR22022}:
\begin{equation}
    \rm L(b\text{\Ha})_{cor}= (17\pm7.6) \times L(b\text{\Ha})_{obs}
\end{equation}
\begin{equation}
    \rm FWHM(b\text{\Ha})_{cor}= (1.92\pm0.22) \times FWHM(b\text{\Ha})_{obs}.
\end{equation}
We estimate the value of $\log ({M_\bullet}/{M_\odot})$ to be $7.6\pm0.2$.
We also utilize the broad Pa{\sc $\beta$} line for the $M_\bullet$ estimation, which is calibrated in \citet{kimNEWESTIMATORSBLACK2010}:
\begin{align}
    M_\bullet &= 10^{7.33\pm0.10}M_\odot \nonumber\\ 
    &\times \left(\frac{L_{\text{Pa{\sc $\beta$},broad}}}{10^{42}{\rm{erg~s^{-1}}}}\right)^{0.45\pm0.03} \left({\rm \frac{FWHM_{\text{{Pa{\sc $\beta$}},broad}}}{\rm 10^3km~s^{-1}}}\right)^{1.69\pm0.16}.
\end{align}
We estimate the value of $\log ({M_\bullet}/{M_\odot})$ to be $6.9\pm0.1$.

We adopt the bolometric luminosity from the SED fitting results to derive the Eddington ratios for different methods by $\lambda_{\rm Edd} = L_{\rm bol}/L_{\rm Edd} = M_{\rm \bullet SED}/M_{\bullet}$, where $M_{\rm \bullet SED}$ is the black hole mass calculated from SED fitting results in 
Table \ref{tab:sed}.

We calculate the black hole mass to the stellar mass ratio, $M_\bullet / M_*$, to be $M_\bullet / M_*=$ 0.02, 0.32, and 0.08 for the black hole masses of the H$\alpha$, H$\alpha$+Sy1.9 correction, and Pa$\beta$ methods, respectively. These ratios fall in the range of $0.02-0.3$ which is comparable with the one of high-$z$ AGN, $\sim 0.001-0.2$, recently reported by JWST studies (e.g., \citealt{harikaneJWSTNIRSpecFirst2023}, \citealt{maiolinoJADESDiversePopulation2023}).
%
%
\begin{table}[tb!]

    \caption{Derived BH properties of CEERS-3506 with different methods}
    \begin{ruledtabular}
    \begin{tabular}{l c c}
     Method & $\log ({M_\bullet}/{M_\odot})$ & $\lambda_{Edd}$\\\hline
    {\Ha} & $6.4\pm0.1$ & 2.06\\
    {\Ha} + Sy1.9 correction & $7.6\pm0.2$ & 0.11\\
    Pa{\sc $\beta$} & $6.9\pm0.1$ & 0.46
    
    \end{tabular}
    \end{ruledtabular}
    \label{tab:BH}
\end{table}


\subsection{Physical Origins of High EW Objects}\label{sec:origin}



\begin{figure*}[!tbh]
    \plotone{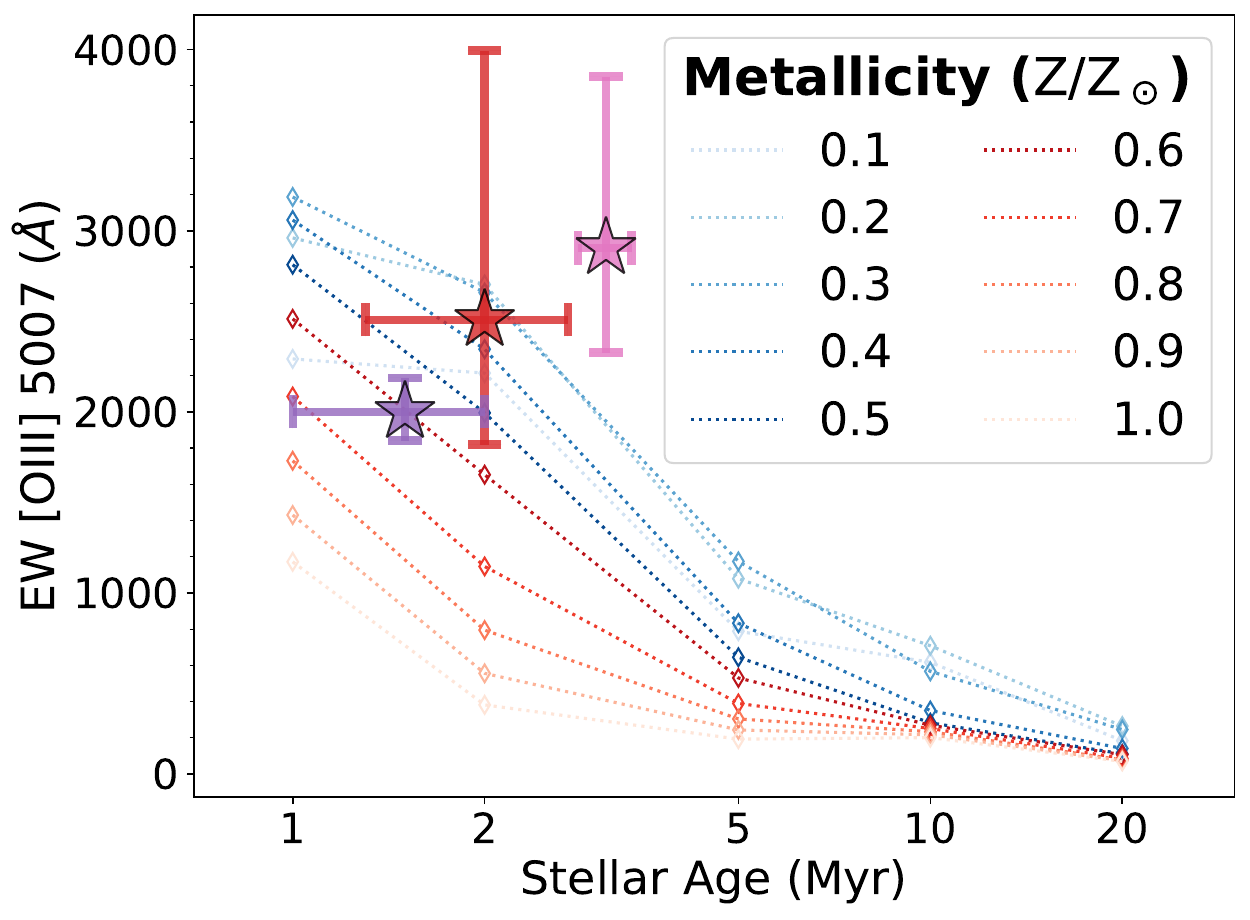}
    \caption{Relation between {\sc EW[Oiii]} and stellar population age for \texttt{BPASS} binary model. Different colors represent various metallicity settings from 0.1 to 1 solar metallicity. The red, pink, and purple stars denote the CEERS-3506, J1000+0211, and J0845$-$0123, respectively. EW({\Oiii}) decreases as the stellar age increases. The relation between EW({\Oiii}) and metallicity is not monotone. Our results agree well with \citet{inoueRestframeUltraviolettoopticalSpectral2011}.}\label{fig:cloudysf}
\end{figure*}
We use \texttt{Cloudy} photoionization code (version C22; \citealt{gunasekera2301Release2023}) to simulate the evolution of EW({\Oiii}). For star formation (SF) models, we utilize the Binary Population and Spectral Synthesis model (\texttt{BPASS} version 2.1; \citealt{eldridgeBinaryPopulationSpectral2017}) with an IMF\footnote{Recommended IMF by BPASS team; \url{https://bpass.auckland.ac.nz/8/files/bpassv2_1_manual_accessible_version.pdf}} upper slope of $-1.3$ and upper stellar mass limit of $100M_\odot$ to generate stellar spectra. We use the models that include binary stars which is more realistic according to current theory and observation. We change the ionizing parameter $\log U$ from $-3.0$ to $-1.0$ by $0.5$ and the stellar age from 1 to 20 Myr. For nebular, we set the electron number density $n_e = 1000~{\rm cm^{-3}} $ and the metallicity $Z=0.1-1.0Z_\odot$ by $0.1Z_\odot$. 

Similarly, we conduct the \texttt{Cloudy} simulation with the ionizing source to be AGN. We adopt the typical AGN spectrum: 
\begin{equation}\label{eq:agn}
    f_\nu = \nu^{\alpha_{uv}}\exp \left(\frac{-h\nu}{KT_{\rm BB}}\right)\exp\left(\frac{-kT_{\rm IR}}{h\nu}\right) + a\nu^{\alpha_x}
\end{equation}
with the Big Bump temperature $T_{\rm BB} = 15,000$K, the low-energy slope of the Big Bump continuum $\alpha_{uv} =0.5$, the slope of the X-ray component $\alpha_x=-1$, the assumed Big Bump infrared exponential cutoff at $kT_{\rm IR} = 0.01{\rm Ryd}$, {and the coefficient $a$ adjusted to produce the X-ray to UV ratio $\alpha_{ox} = -1.4$ for the case where the Big Bump does not contribute to the emission at $\rm 2~Kev$. }

\begin{figure*}[!hbt]
    \plotone{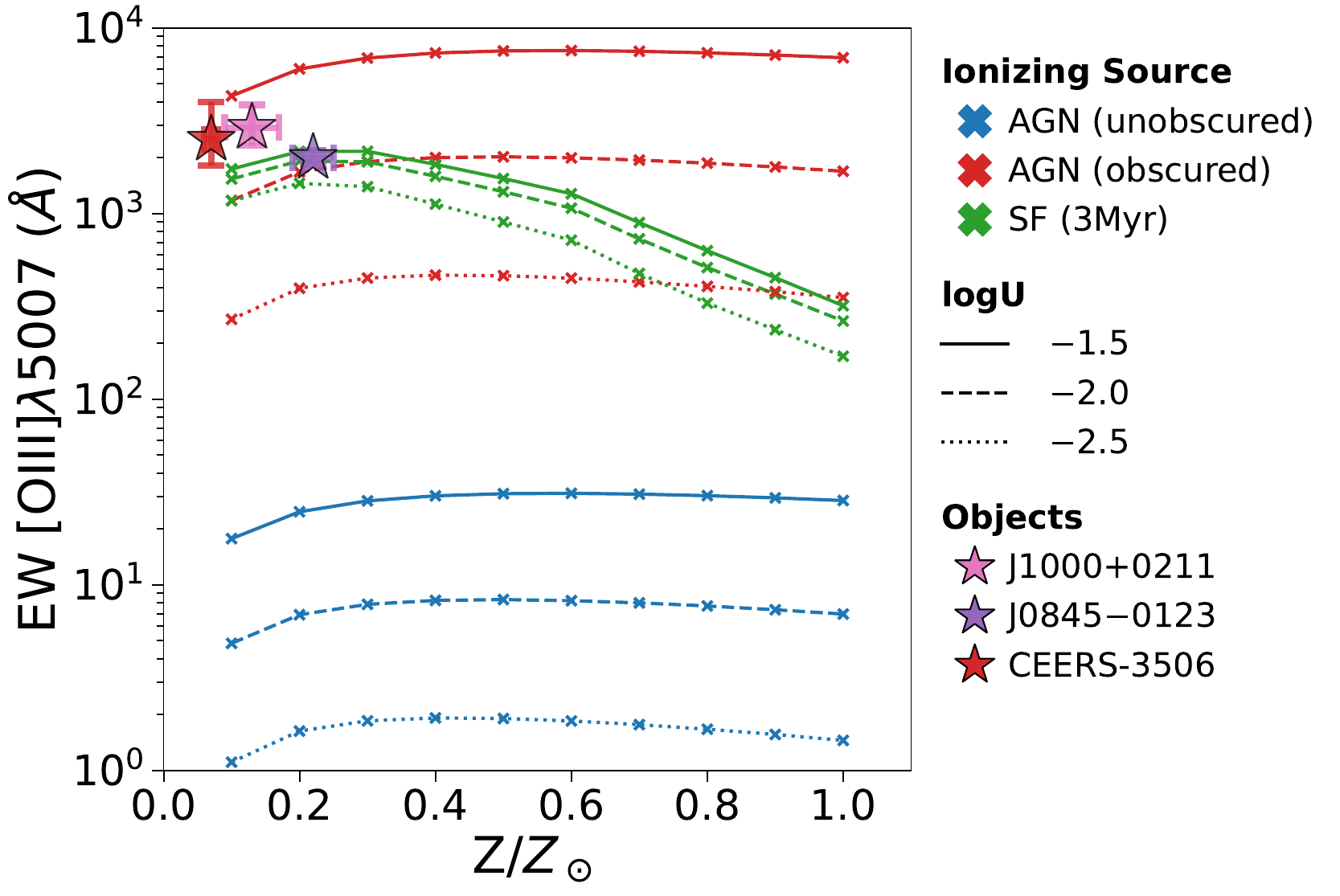}
    \caption{Relation between EW({\Oiii}) and metallicity for different models. For clarity, we present our simulation results with representative parameters. The green, red, and blue lines denote the EWs for star formation models of BPASS binary models at the stellar age of 3 Myr, obscured AGN, and unobscured AGN, respectively. The solid, dashed, and dotted lines denote the EWs for the ionizing parameter ($\log U$) of $-1.5$, $-2.0$, and $-2.5$. The stars mark our objects.}
    \label{fig:EW_Z}
\end{figure*}

In Figure \ref{fig:cloudysf}, we show the evolution of EW({\Oiii}) as stellar age increases for the SF models. We find that the two objects, CEERS-3506 and J0845$-$0123, can agree with the SF models, while it is hard to explain high EW({\Oiii}) of J1000+0211 by the SF models. In Figure \ref{fig:EW_Z}, we show the relation between EW({\Oiii}) and metallicity for the SF and AGN models. We find that the obscured AGN models can produce high EW({\Oiii}) when the ionizing parameter is high ($\log U>-2$). Furthermore, we compare the observables of the three {\Oiii} emitters with our models. We find that the obscured AGN model, whose host galaxies have a weak stellar continuum at rest-frame optical, can produce the high EW({\Oiii}) of our sources.

{In addition, J1000+0211 and CEERS-3506 lie within the deep Chandra fields, specifically within the Chandra COSMOS survey \citep{Civano_2016, Marchesi_2016} and the AEGIS-X Deep survey, respectively. However, neither source is listed in the corresponding X-ray catalogs, with the absence of X-ray detection potentially indicating significant obscuration.}

\section{Summary}\label{sec:sum}
In this paper, we report the physical properties of three EELGs with strong {\OOiii} emission lines. The pronounced EWs of their emission lines couple with a faint continuum, and thus necessitate sensitive observations for reliable detection of continuum emission. 
Below, we summarize our principal findings:

\begin{itemize}
    \item We present deep Subaru/FOCAS VPH850 spectra of two most extreme {\Oiii} emitter candidates. We estimate EWs({\Oiii}) with detected continnua ($2000^{+188}_{-159}${\AA} for J0845$-$0123; $2905^{+946}_{-578}${\AA} for J1000+0211; Table \ref{tab:line}).  We find another extreme {\Oiii} emitters in CEERS program with EW({\Oiii}) $=2508^{+1487}_{-689}${\AA}.
    \item Despite the absence of clear AGN signatures from optical line diagnostics, our analysis reveals strong near-infrared excess in the SEDs of two galaxies, indicative of obscured AGN activity. The detection of broad {\Ha}, He{\sc i$\lambda$10830}, and Pa{\sc $\beta$} in the CEERS-3506 further support of AGN activity, allowing us to estimate black hole masses and explore the black hole-to-stellar mass relationship. Using various calibration methods, we estimate $\log (M_\bullet/M_\odot)\sim 6.4-7.6$, corresponding to a black hole to stellar mass ratio of $M_\bullet/M_*\sim0.02-0.32$.
    \item We measured the metallicity of the three objects with $Z\sim0.07-0.20Z_\odot$ using the direct temperature method with {\sc [Oiii]$\lambda$4363}. To interpret our findings, we employ \texttt{Cloudy} photoionization models, considering a range of parameters including stellar and AGN incident spectra, metallicities, and ionization parameters. Our models indicate that the large EWs({\Oiii}) cannot be fully explained by stellar or unobscured AGN spectra alone, but are more consistent with the presence of obscured AGN. The models successfully reproduce the observed EWs({\Oiii}) by invoking a scenario where ionizing photons are efficiently produced by obscured AGN with weak nuclear and stellar continua, matching the SED shapes.
    
    \item We propose that within the EELG population, particularly those with the most extreme EWs({\Oiii}), the fraction of AGN is likely higher than what optical line diagnostics suggest. However, larger sample sizes are needed to confirm this implication. This requires revisiting previously reported high-EW({\Oiii}) emitters (e.g., Blueberry galaxies from \citealt{yangBlueberryGalaxiesLowest2017}, other {\Oiii} emitters from \citealt{matsuokaSubaruHighExploration2018, matsuokaSubaruHighzExploration2019}) with JWST near-infrared and mid-infrared data.
\end{itemize}


\section{Acknowlegements}
We appreciate Ding Xuheng's suggestions about the morphology analysis. We acknowledge Aya Bamba, Takao Nakagawa, and Crystal Martin for their discussion about this work. 
This publication is based upon work supported
by the World Premier International Research Center
Initiative (WPI Initiative), MEXT, Japan, KAKENHI
(20H00180, 24H00245, and 21K13953) through Japan
Society for the Promotion of Science. 
KI acknowledges support under the grant PID2022-136827NB-C44 provided
by MCIN/AEI/10.13039/501100011033 / FEDER, UE.
This work was supported by
the joint research program of the Institute for Cosmic
Ray Research (ICRR), University of Tokyo.

This research is based in part on data collected at the Subaru Telescope, which is operated by the National Astronomical Observatory of Japan.  
The observations were carried out within the framework of Subaru-Keck time exchange program which is operated by the National Astronomical Observatory of Japan. We are honored and grateful for the opportunity of observing the Universe from Maunakea, which has the cultural, historical and natural significance in Hawaii.
Some of the data presented herein were obtained at Keck Observatory, which is a private 501(c)3 non-profit organization operated as a scientific partnership among the California Institute of Technology, the University of California, and the National Aeronautics and Space Administration. The Observatory was made possible by the generous financial support of the W. M. Keck Foundation.
We are honored and grateful for the opportunity of observing the Universe from Maunakea, which has the cultural, historical, and natural significance in Hawaii.
Some of the data products presented herein were retrieved from the Dawn JWST Archive (DJA). DJA is an initiative of the Cosmic Dawn Center (DAWN), which is funded by the Danish National Research Foundation under grant DNRF140.

\bibliography{ref}



\end{document}